\documentclass[
 reprint,
 amsmath,amssymb,
 aps,
]{revtex4-1}

\usepackage{graphicx}
\usepackage{dcolumn}
\usepackage{bm}
\usepackage{hyperref}

\usepackage[dvipsnames]{xcolor}

\begin{document}

\preprint{APS/123-QED}

\title{Dilepton production and resonance properties within a new hadronic transport approach in the context of the GSI-HADES experimental data}

\author{Jan Staudenmaier$^{1,2}$, Janus Weil$^1$, Vinzent Steinberg$^1$, Stephan Endres$^1$, Hannah Petersen$^{1,2,3}$}

\address{$^1$ Frankfurt Institute for Advanced Studies (FIAS), Ruth-Moufang-Stra{\ss}e 1, 60438 Frankfurt am Main, Germany}
\address{$^2$ Institut f\"ur Theoretische Physik, Johann Wolfgang Goethe-Universit\"at, Max-von-Laue-Str. 1, 60438 Frankfurt am Main, Germany}
\address{$^3$ GSI Helmholtzzentrum f\"ur Schwerionenforschung, Planckstr. 1, 64291 Darmstadt, Germany}

\date{\today}

\begin{abstract}
The dilepton emission in heavy-ion reactions at low beam energies is examined within a hadronic transport approach. In this article, the production of electron-positron pairs from a new approach named SMASH (Simulating Many Accelerated Strongly-interacting Hadrons) is introduced. The dilepton emission below the hadronic invariant mass threshold is taken into account for all direct vector meson decays. The calculations are systematically confronted with HADES data in the kinetic-energy range of $1 - 3.5A$ GeV for elementary, proton-nucleus and nucleus-nucleus reactions. The present approach employing a resonance treatment based on vacuum properties is validated by an excellent agreement with experimental data up to system sizes of carbon-carbon collisions. After establishing this well-understood baseline in elementary and small systems, medium effects are investigated with a coarse-graining approach based on the same hadronic evolution. The effect of in-medium modifications to the vector meson spectral functions is important for dilepton invariant mass spectra in ArKCl and larger systems, even though the transport approach with vacuum properties reveals similar features due the coupling to baryonic resonances and the intrinsically included collisional broadening. This article provides a comprehensive comparison of our calculations with published dielectron results from the HADES collaboration. In addition, the emission of dileptons in gold-gold and pion-beam experiments, for which results are expected, is predicted.
\end{abstract}

\pacs{Valid PACS appear here}
\maketitle

\section{Introduction}

Dileptons are a clean probe for hot and dense matter, as it is studied in heavy-ion collisions in a wide range of beam energies. Since they only interact electromagnetically, they escape the medium nearly unperturbed, thus allowing unique access to the properties of the medium and resonances that decay within a strongly interacting medium. In contrast, hadronic decay products suffer from re-scattering and absorption such that the interesting information about the hot and dense stage of the reaction is masked.

More specifically, dilepton production offers a complementary perspective on the spectral function of the vector mesons, which is reflected in the invariant mass spectrum of the lepton pairs. A modification of these spectral functions might indicate the restoration of chiral symmetry \cite{Hatsuda:1991ez,Pisarski:1981mq,Brown:1991kk,Rapp:2004zh}. Chiral symmetry is of major interest to research, since its breaking accounts to a large extent for the mass generation of the visible matter~\cite{Koch:1997ei}. The corresponding modifications to the spectral function of vector mesons inside a hot and dense medium have been discussed in the literature \cite{Rapp:2009yu,Leupold:2009kz,Hayano:2008vn}. Two modification scenarios were prevalent, in particular for the $\rho$ meson: a shift in mass or a broadening of the spectral functions~\cite{Brown:2003ee,Rapp:1999us}. For example, calculations based on hadronic many-body theory \cite{Rapp:1999us} and the functional renormalization group \cite{Jung:2016yxl} are performed to obtain a quantitative understanding of this effect.

This work focuses on the dilepton production in elementary, nucleon-nucleus and nucleus-nucleus collisions. Experimentally, the emission of dielectrons or dimuons was studied at a number of different facilities. At CERN SPS the high-quality experimental data of dimuon production in indium-indium collisions recorded by NA60~\cite{Damjanovic:2005ni, Arnaldi:2006jq} allowed the investigation of the $\rho$ spectral function and confirmed the previous dilepton measurement from CERES~\cite{Agakishiev:1995xb} that revealed an excess in the low invariant mass region. The findings of NA60 essentially settled the debate of how the $\rho$ spectral shape changes inside a hot and dense medium. The data are consistent with a broadening and disfavor a mass shift of the $\rho$~\cite{vanHees:2006ng}. For higher energies up to $\sqrt{s_{NN}}=200$ GeV, dileptons are measured at RHIC by STAR~\cite{Adamczyk:2013caa} and PHENIX~\cite{Adare:2015ila}. Both also report an enhancement in the dilepton invariant mass range from $0.30$ to $0.76$ GeV that is again attributed to a broadening of the $\rho$ spectral function. The present work focuses on the dielectron production in the kinematic regime of beam energies of $E_{\rm{Kin}}=1-3.5A\,\textrm{GeV}$, which is covered by the HADES experiment~\cite{Agakishiev:2009yf,Agakishiev:2012tc,HADES:2011ab,Agakishiev:2012vj,Agakishiev:2007ts,Agakichiev:2006tg,Agakishiev:2011vf} at the GSI facility. The HADES results confirmed previous measurements from the DLS collaboration~\cite{Porter:1997rc}. In the future, the CBM experiment at FAIR~\cite{Friman:2011zz} will add to the existing experimental data with new results from the intermediate energy range, specifically probing,  together with complementary programs from NICA and J-PARC, the dilepton emission from the high net baryon density region.

To connect the theoretical calculations of the vector meson spectral functions with experimental measurements, dynamical approaches that describe the full evolution of heavy-ion collisions in detail have to be employed. Hadronic transport approaches are applied successfully \cite{Bass:1998ca, Nara:1999dz, Bratkovskaya:2011wp, Buss:2011mx} for low beam energy collisions. In the current work, a new hadronic transport approach, SMASH (Simulating Many Accelerated Strongly-interacting Hadrons)~\cite{Weil:2016zrk}, is introduced.

SMASH combines and adapts different successful aspects of previous transport approaches. For the here discussed dilepton production the resonance description aspects are most relevant. The resonance states employed in UrQMD~\cite{Bass:1998ca} are used as the foundation. Their properties e.g. their branching ratios are constrained with PDG data~\cite{Agashe:2014kda} and experimental dilepton and cross section data. Compared with the published results for the dilepton production in UrQMD~\cite{Schmidt:2008hm} more recent and additional data is used. The spectral function is also treated as in UrQMD, but the decay width treatment is the same as in GiBUU~\cite{Buss:2011mx}. However, no \emph{off-shell propagation} is taken into account like in the HSD~\cite{Bratkovskaya:2011wp} or the GiBUU approach. Details specific to the dilepton production are outlined below. Comparisons in this work are focused on GiBUU and UrQMD, since they are most comparable to the SMASH approach. For a recent, more general comparison of the different approaches, the reader is referred to~\cite{Zhang:2017esm}.

Dilepton emission within hadronic transport approaches has been extensively explored by previous work using the GiBUU~\cite{Weil:2012ji}, the HSD~\cite{Bratkovskaya:2013vx} and the UrQMD~\cite{Schmidt:2008hm} approach. These studies cover a variety of aspects concerning the dilepton production at low energies such as the effect of the coupling of the $\rho$ meson to baryonic resonances, Bremsstrahlung, the $\Delta$ contribution and the density dependence. The dilepton emission is theoretically based to a large extent on the GiBUU approach (see Sec.~\ref{sec:dil_prod}), since it has recently proven to be successful in describing the experimental data~\cite{Weil:2012ji}.
  
The previous efforts establish a solid foundation for the work presented here, which nevertheless allows us to add two new aspects: First, low-mass contributions to the vector meson decay channels are studied for all vector mesons. The spectral function for vector mesons does not vanish at the hadronic threshold. Instead, it vanishes at $2m_e$, the smallest possible invariant mass  of the decay products, when including $V\rightarrow e^+e^-$ decays ($V=\rho,\omega,\phi$). The treatment leads consequently to low-mass contributions below the hadronic threshold for direct vector meson dilepton decays. These low-mass dilepton yields are investigated. In particular, their significance relative to other decay channels and for the total yield is studied. In this work, all vector mesons ($\rho,\omega,\phi$) are included. This is an extension of previous work with the GiBUU approach, which considered such contributions only for the $\rho$ meson. The UrQMD approach, on the other hand, neglects them entirely for numerical reasons. 
 
Secondly, the employment of a coarse-graining approach~\cite{Endres:2015fna} in combination with the just discussed low-mass contributions offers the unique opportunity of a direct comparison (see Sec.~\ref{sec:cg_comp}). It is possible to investigate different medium effects on the dilepton spectrum including the low-mass region based on the same hadronic evolution. The coarse-graining approach employs in-medium spectral functions for the vector mesons whereas the transport calculation's spectral functions are based on vacuum properties. Therefore, a comparison of the decay yield of the vector mesons produced by both approaches is of particular interest.
Although results for a coarse-grained UrQMD dilepton production are reported~\cite{Endres:2015fna,Galatyuk:2015pkq}, such a direct comparison has not been performed. Additionally, the low-mass contributions are neglected for the UrQMD transport calculations, making a comparison to the corresponding coarse-grained results in this mass region unfeasible. For GiBUU no coarse-grained results are reported so far.

In addition, to these new results, this work is motivated by the validation of and constraints for the SMASH approach in general and the resonance description in particular. They are provided by the calculation of the dilepton yield and the subsequent comparison to experimental data. It is for example possible to specifically probe and constrain the branching ratios of decays of baryonic resonances into vector mesons. The dilepton production furthermore facilitates comparisons to other approaches as well as predictions for newly measured collision systems by HADES~\cite{Scozzi:2017sho, Franco:2017ano}. A well controlled baseline of the dilepton production for the hadronic sector with vacuum properties is also essential to study higher energies e.g. in a hybrid approach~\cite{Santini:2009nd,Petersen:2014yqa}. There the final state non-equilibrium emission is calculated with hadronic transport approaches. Studying the effect of late stage re-scattering on the dilepton production is planned for future work. Emission of dileptons also remains a key observable for upcoming facilities focusing on high-density collisions, e.g. CBM at FAIR, which will also be explored with the SMASH approach in the future.

As discussed at the beginning of this section, medium effects are closely related to the study of dileptons and are therefore a focus of the discussion of the results reported here. The microscopic transport includes the so called \emph{collisional broadening}. Since absorption of resonances by other hadrons is dynamically accounted for, the lifetime is shortened in the presence of a hadronic medium. The reduction of the lifetime equals an effective broadening of the decay width of the resonance. The later employed coarse-graining approach additionally includes vector meson spectral functions that are explicitly dependent on the temperature and density of the medium~\cite{Rapp:1999us, Rapp:2000pe}. Such modifications will be labeled as \emph{in-medium modifications} in the following, since they are only employed for the evolution within a hot and dense medium. In this context, the coupling of baryons to the vector mesons is of particular interest~\cite{Rapp:2009yu}. In SMASH this coupling is included as the decay of baryonic resonances into vector mesons. Whereas in the calculation of the in-medium spectral functions, the coupling enters as an important contribution to the self-energy. This is done in a self-consistent way including interference terms. On the other hand, the presented transport calculation neglects interference terms as well as broadening of the spectral function originating from a consistent treatment of the self-energies in the vector meson propagator as an approximation. The results reported in the following allow us to access these medium effects, since they affect the dilepton invariant mass spectra significantly. In particular, the mentioned direct comparison between the emission of the transport and the coarse-graining approach allows us to contrast the different employed effects.

This work is outlined as follows: First, in Sec.~\ref{sec:desc} the SMASH approach is introduced with an emphasis on the employed resonance and dilepton production. The results in Sec.~\ref{sec:res} are sorted by system size, where the studied systems are chosen according to the HADES program. As a start, the dilepton production in elementary reactions in Sec.~\ref{sec:res_elem} and the cold nuclear matter scenario of proton-nucleus reaction in Sec.~\ref{sec:res_pA} are calculated in order to establish the above mentioned baseline for the lepton pair emission. Afterwards, nucleus-nucleus collisions are studied in Sec.~\ref{sec:res_AA}. For the large systems the coarse graining approach based on the same hadronic evolution is explored (Sec.~\ref{sec:cg}), which enables us to probe the sensitivity to in-medium modifications of the vector meson spectral functions in larger systems. This additionally permits the direct comparison mentioned above between the dilepton emission based on a vacuum and in-medium description of resonances. Finally, a summary and outlook of the presented work is supplied in Sec.~\ref{sec:sum}.

\section{Model Description} \label{sec:desc}

\subsection{SMASH}

The approach applied in this work is SMASH (Simulating Many Accelerated Strongly-interacting Hadrons), which is a new hadronic transport approach for the dynamical description of collisions at low and intermediate beam energies and dilute non-equilibrium stages of heavy-ion collisions. The goal is to provide a standard reference for hadronic systems with vacuum properties. SMASH is a microscopic approach based on the relativistic Boltzmann equation. The collision term is modelled by excitation and decay of resonances and is restricted to binary collisions. Two particles collide, if the geometric collision criterion ($d_{\rm{trans}}<\sqrt{\sigma_{\rm{tot}}/\pi}$) is fulfilled. Only binary collisions and two body decays are performed in order to conserve detailed balance. Generic multi-particle decays (e.g. $\omega\rightarrow3\pi$) are incorporated by assuming intermediate resonance states ($\omega\rightarrow\rho\pi\rightarrow3\pi$). The model includes all well known hadronic degrees of freedom listed by the PDG~\cite{Agashe:2014kda} up to a mass of $2.35$ GeV. The agreement with elementary cross section data up to 4 GeV as well as reasonable agreement for proton and pion spectra with experimental data for $E_{\rm{Kin}} = 1-2A\,$GeV is shown in \cite{Weil:2016zrk}, which also includes a comprehensive description of the approach. The version used for this work is SMASH-1.1.

\subsubsection{Resonance description \label{res_desc}}

Since dileptons are sensitive to resonance properties, the treatment of resonances is of great importance. In general, the current work mainly concentrates on vacuum spectral functions to provide a baseline for additional in-medium modifications.

All resonance spectral functions are relativistic Breit-Wigner functions:

\begin{equation}
\label{eq:spectral}
\mathcal{A}(m) = \frac{2\mathcal{N}}{\pi} \frac{m^2 \Gamma(m)}{(m^2 - M_0^2)^2 +
^2 \Gamma(m)^2}\,.
\end{equation}
Here $M_0$ is the pole mass and $m$ is the actual mass of the resonance. The normalization factor $\mathcal{N}$ is chosen such that

\begin{equation}
\int\displaylimits_0^\infty dm\mathcal{A}(m) = 1\,.
\end{equation}

It deviates from 1, since the decay width is mass-dependent. The deviation is below 55\% for all particles, with most normalization factors close to unity.

Energy-momentum conservation is always enforced in the propagation and creation of resonances. In this sense, all resonances are always on the mass shell (\emph{on-shell}).

\begin{figure}
\includegraphics[width=0.95\columnwidth]{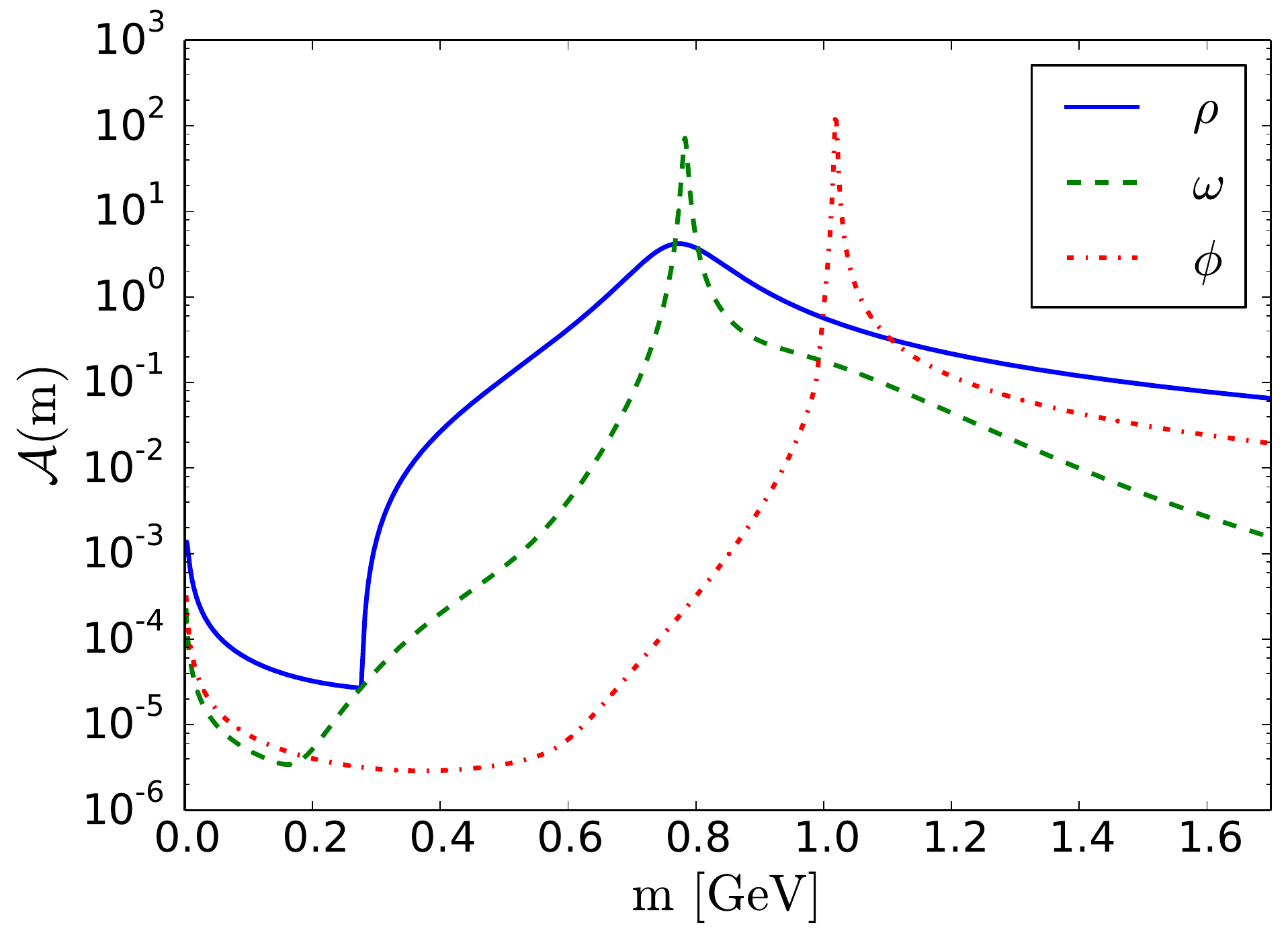}\caption{\label{fig:spectral}Spectral Function of the vector mesons with dielectron decay mode.}
\end{figure}

The spectral functions vanish at the combined mass of the lightest decay products. It is important to note that the dilepton decay mode is correctly taken into account as the kinematic threshold. The spectral functions of resonances that directly decay into a lepton pair ($R\rightarrow l^+l^-$) have contributions below the lightest combined mass of hadronic decay products, which is referred to as the hadronic threshold here. In Fig.~\ref{fig:spectral} the spectral function of the three vector mesons with dielectron decay channel are depicted as an example. The spectral functions peak at the pole mass $M_{0}$ and, as expected from their widths, the $\rho$ peak is the broadest, followed by the $\omega$ and the sharp $\phi$. Noticeable is the kink for the $\rho$ spectral function at around $0.3$ GeV. The decay with the lightest hadronic decay products for the $\rho$ is $\rho \rightarrow \pi^+\pi^-$, so the hadronic threshold is at $2m_\pi$. This threshold leads to the kink, because the partial hadronic decay width of the $\pi$ decay vanishes at this mass. Since the $\rho$ can also decay into an $e^+e^-$ pair the spectral function continues down to $2m_e$. This is also true for the $\omega$ and the $\phi$. Both spectral functions also have contributions down to the combined mass of the actual lightest decay products -- the dielectron mass.

This treatment of thresholds was introduced in \cite{Weil:2012ji} for the $\rho$ meson and marks a difference to other approaches \cite{Bratkovskaya:2007jk, Schmidt:2008hm}, which neglect the contributions below the hadronic threshold for numerical reasons. To study the significance of those sub-threshold dilepton contributions and to be consistent, the treatment is extended to all three vector mesons in this work. We find that they affect the dilepton spectrum in the low-mass region (see the discussion in~\ref{sec:pp}, \ref{sec:CC} and \ref{sec:auau}). In a Monte-Carlo approach it is however challenging to numerically populate the mass region below the hadronic threshold, since the spectral function (Fig.~\ref{fig:spectral}) in this region only has small values. This leads to visible statistical fluctuations in some of the following dielectron invariant mass spectra.

The non-trivial shape of the spectral function, as e.g. seen in Fig.~\ref{fig:spectral}, originates from the mass dependent decay width $\Gamma(m)$, which is the sum of all partial decay widths for the different decay modes.

\begin{equation}
\Gamma(m)=\sum_i \Gamma_i(m)
\end{equation}

The lifetime of the resonances is given as $\tau = 1/\Gamma (m)$. We have verified that changing the lifetime to be $\tau = 1/\Gamma (M_0)$ does not alter results significantly in elementary reactions. All hadronic partial widths are calculated following the framework of Manley et al. \cite{Manley:1992yb} (without taking the exact same parameters for the resonance properties). The partial width of a two-body resonance decay $R\rightarrow ab$ is calculated as follows

\begin{equation}
\Gamma_{R\rightarrow ab} = \Gamma^0_{R\rightarrow ab} \frac{\rho_{ab}(m)}{\rho_{ab}(M_0)}.
\end{equation}
The function $\rho_{ab}(m)$ is defined as

\begin{align}
\label{eq:rho}
       \rho_{ab}(m) = \int & dm_a dm_b \mathcal{A}_a(m_a)\mathcal{A}_b(m_b) \nonumber \\
             \times & \frac{|\vec{p}_f|}{m} B_L^2(|\vec{p}_f|R) \mathcal{F}_{ab}^2(m)\,,
\end{align}
where $m_a$ and $m_b$ are the masses of the decay products $a$ and $b$, $\mathcal{A}_{a/b}$ is their respective spectral function and $|\vec{p}_f|$ the absolute value of the final-state momentum of $a$ and $b$ in the center-of-momentum frame. Equation~\ref{eq:rho} also includes the "Blatt-Weisskopf functions" $B_L$ \cite{BlaWei} and the form factor $\mathcal{F}_{ab}$. For a more detailed description of the resonance treatment the reader is referred to \cite{Weil:2016zrk}. The decay widths used for dilepton decays are described in section~\ref{sec:dw_dil}.

\subsubsection{Elementary cross sections \label{sec:xs}}

\begin{figure}
\includegraphics[width=0.95\columnwidth]{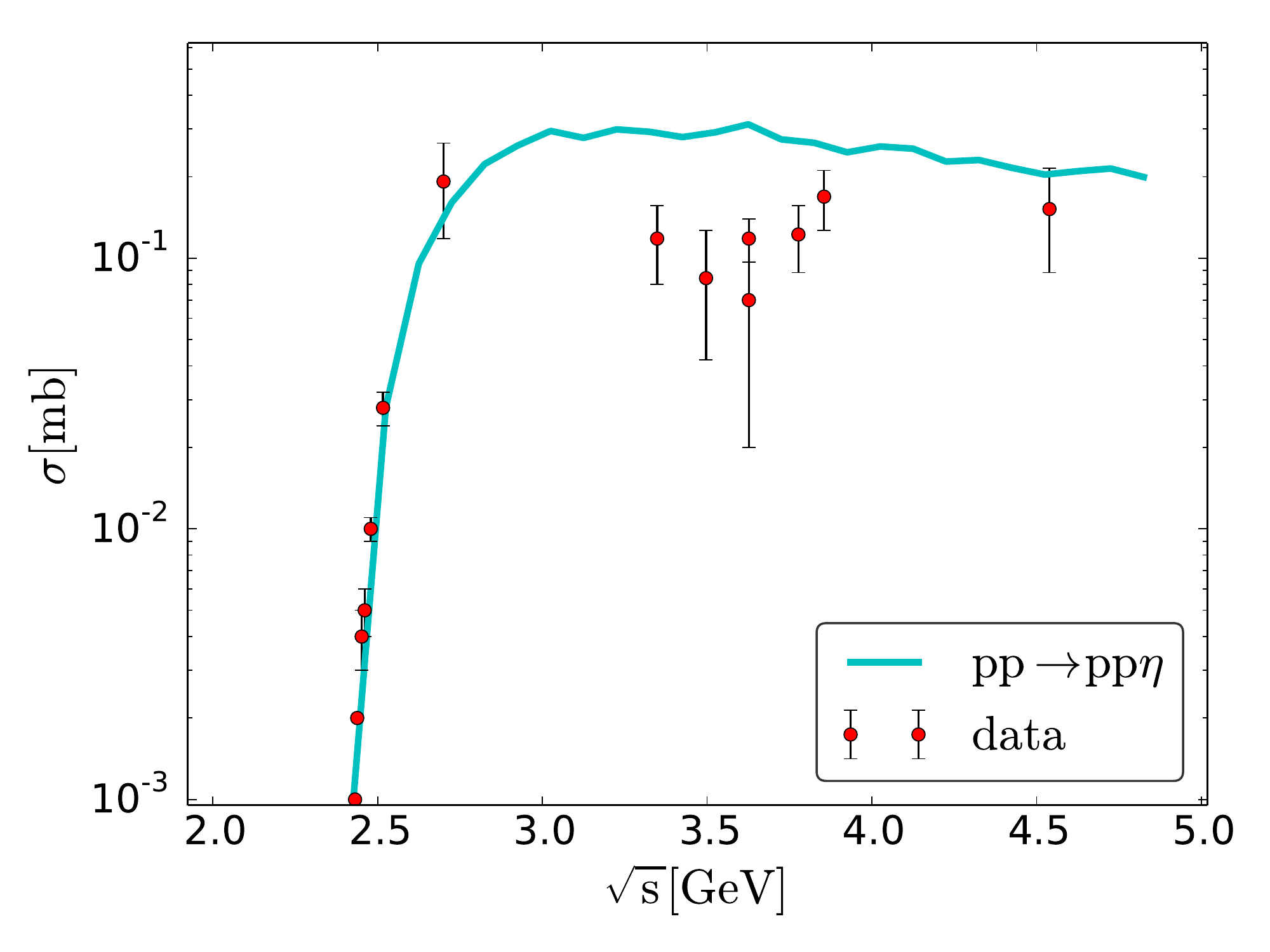}
\caption{\label{fig:xs_eta_pp} Production cross section for $pp\rightarrow pp\eta$. Experimental data from \cite{Calen:1997sf,CHIAVASSA1994270,LB}.}

\includegraphics[width=0.95\columnwidth]{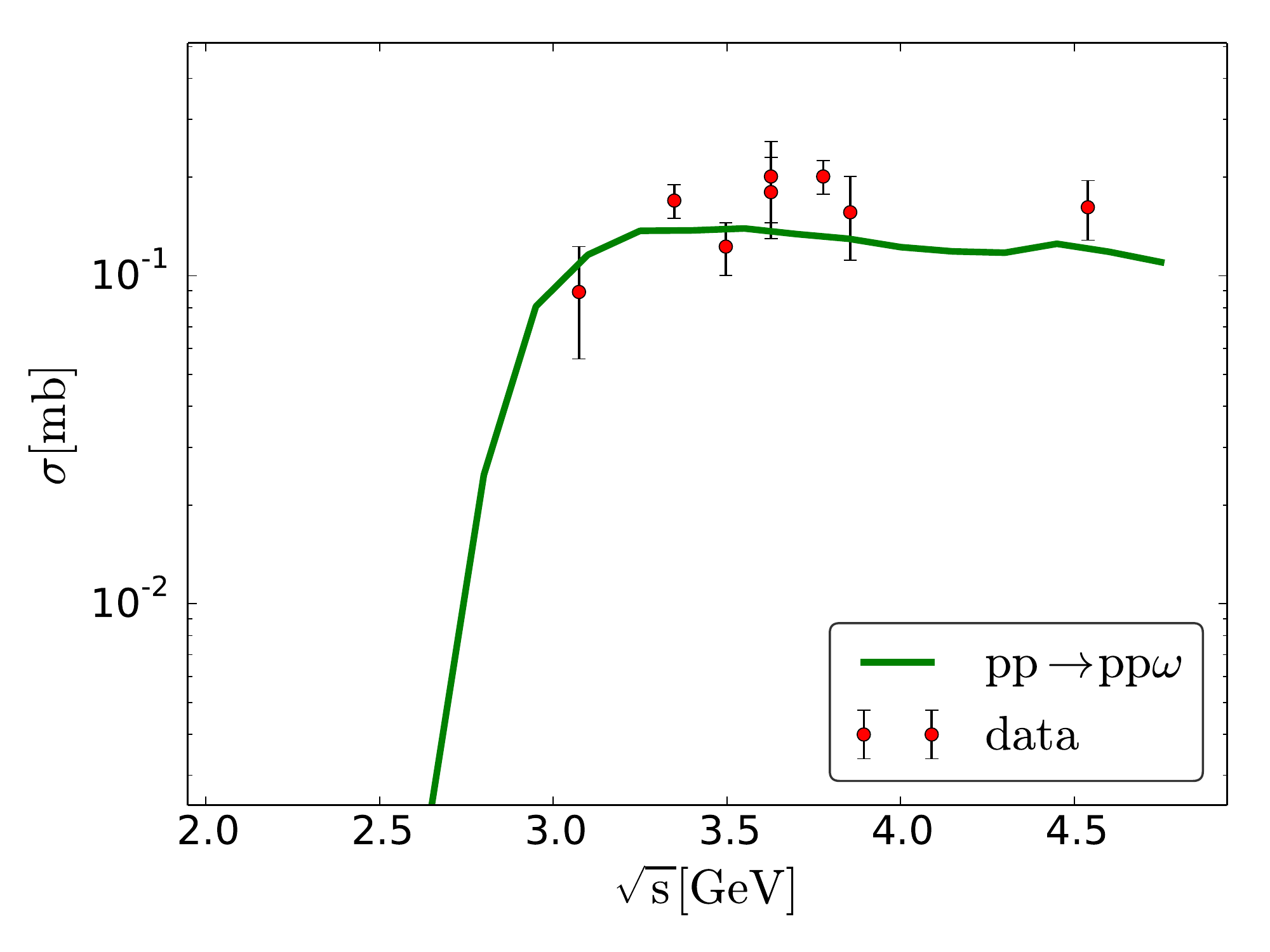}
\caption{\label{fig:xs_omega} Production cross section for $pp\rightarrow pp\omega$. Experimental data from \cite{LB}.}

\includegraphics[width=0.95\columnwidth]{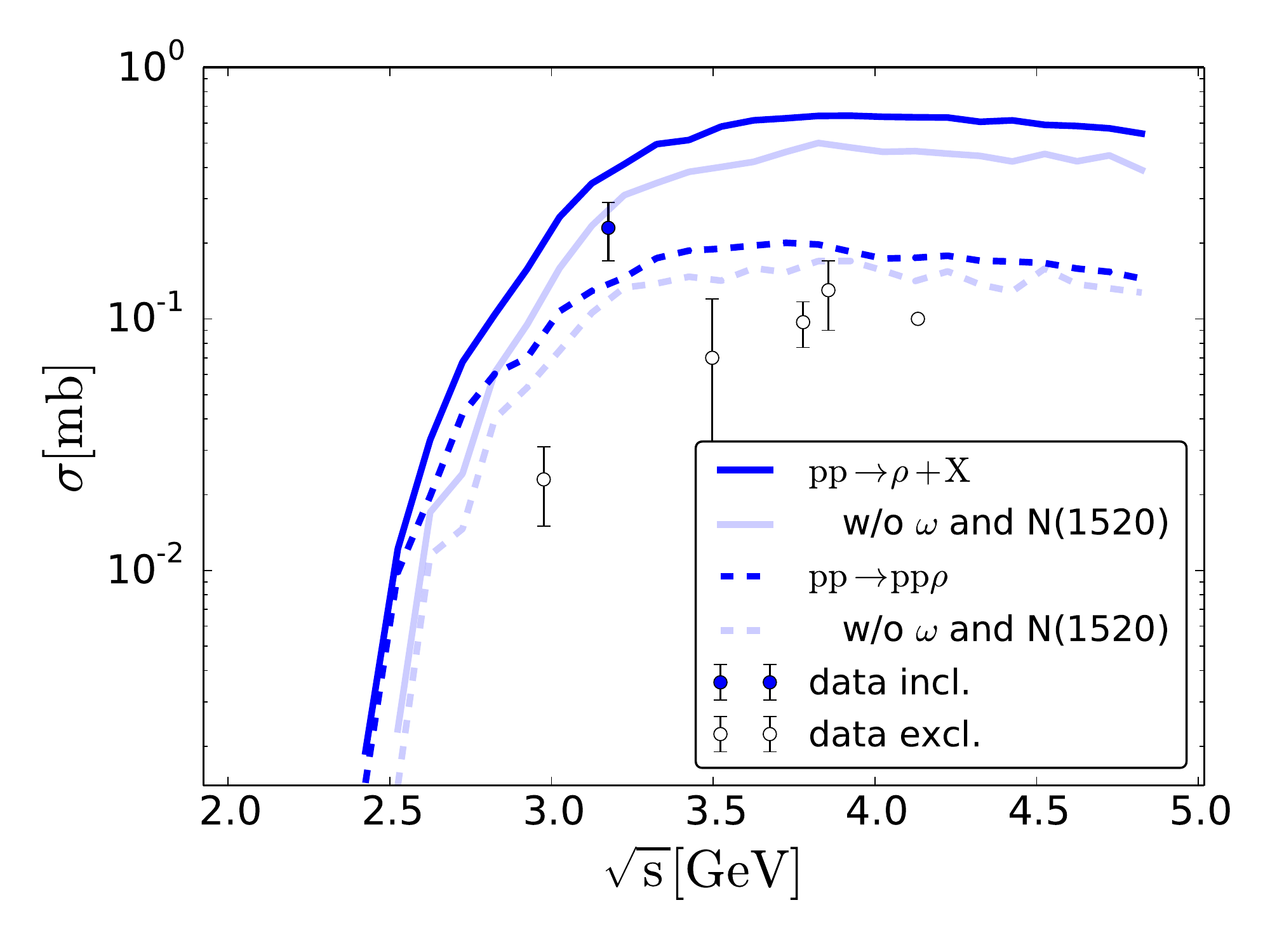}
\caption{\label{fig:xs_rho} Exclusive production cross section for $pp\rightarrow pp\rho$ (empty data points from \cite{Flaminio:1984gr,LB}) and inclusive $pp\rightarrow\rho + X$ cross section (full data point from \cite{HADES:2011ab}).}
\end{figure}

Cross section data are in general a valuable tool to constrain the particle production for different collision energies. This section includes several results for particles that decay into dileptons. Their production therefore directly influences the dilepton production studied in this work. The cross section results complement the already reported total and single pion production cross sections in \cite{Weil:2016zrk}. Beginning with the $\eta$ production cross section in pp collisions, Fig.~\ref{fig:xs_eta_pp} shows the exclusive $\eta$ production in $pp\rightarrow pp\eta$. A good agreement is observed close to the threshold, whereas too many $\eta$ mesons are produced from $pp\rightarrow pp\eta$ for $\sqrt{s} > 3.25$ GeV. The disagreement however does not affect the few-GeV energy range of HADES measurements studied in this work. Error bars in Fig.~\ref{fig:xs_omega}, which shows the exclusive $\omega$ production cross section are large and SMASH results are in reasonable agreement with them. In the case of the $\rho$ meson, both inclusive ($pp\rightarrow\rho + X$) and exclusive cross sections ($pp\rightarrow pp\rho$) are shown (Fig.~\ref{fig:xs_rho}). Both are overestimated (solid lines) compared to the experimental data points, especially the exclusive cross section, which dominates the inclusive cross section for energies close to the threshold. It is important to consider here that the $\rho$ meson in SMASH is used as an intermediate state to emulate different Dalitz decays in two steps following the idea of strict Vector Meson Dominance~\cite{OConnell:1995nse}. The advantage of this treatment is the conservation of detailed balance. The two most prominent decays are the $3\pi$ decay of the $\omega$ ($\omega\rightarrow\rho\pi\rightarrow 3\pi$) and the $N^*(1520)$ dilepton Dalitz decay, which is emulated by $N^*(1520)\rightarrow\rho N\rightarrow e^+e^-N$. Neglecting these additional proxy contributions (transparent lines in Fig.~\ref{fig:xs_rho}) leads to an agreement within errors for the inclusive cross section data point, but not for the exclusive channel. The overestimation in the inclusive $\rho$ production therefore can be seen as a compromise of the two-step treatment of three-body decays.

\subsubsection{Isospin-asymmetric $\eta$ production}
\begin{figure}
\includegraphics[width=0.95\columnwidth]{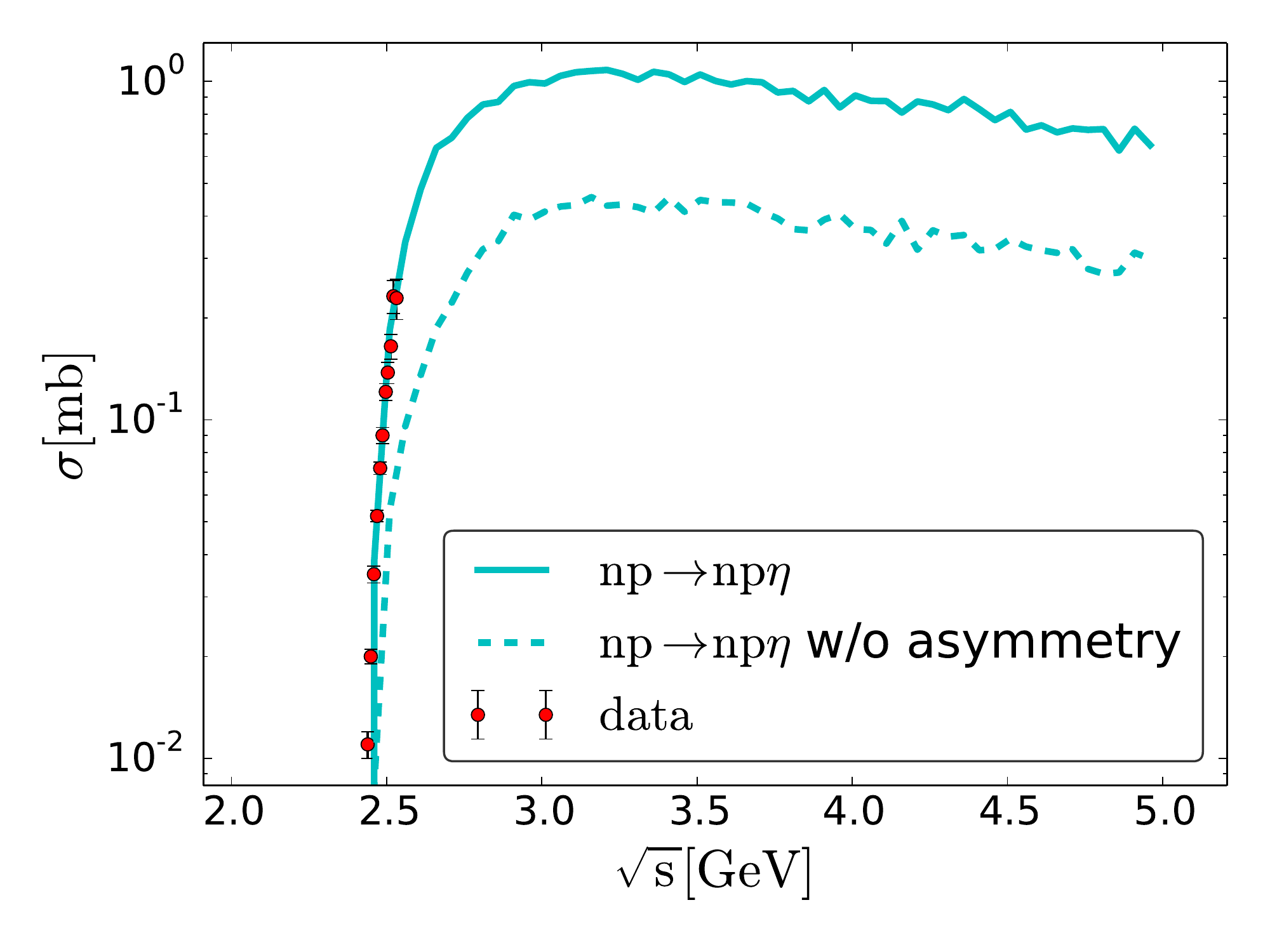}
\caption{\label{fig:eta_prod_top} Cross section for $pn\rightarrow pn\eta$. Experimental data from \cite{Calen:1998vh}.}
\end{figure}

In general SMASH assumes isospin symmetry in the production of particles, but there are a few exceptions of this treatment for NN reactions. One is newly introduced for this work to improve the description of the dilepton yield stemming from the Dalitz decay of the $\eta$ meson. There exists experimental evidence that the $\eta$ production via $pn\rightarrow pn\eta$ is enhanced in comparison to $pp\rightarrow pp\eta$ by approximately a factor of 6.5~\cite{Calen:1998vh}. The dominant source of the $\eta$ meson for low energies is the decay of $N^*(1535)$, therefore we enhance its production following the suggestion from \cite{Teis:1996kx} by modifying the matrix element,
\begin{equation}
|\mathcal{M}_{pn\rightarrow NN^*(1535)}|^2=6.5\times|\mathcal{M}_{pp\rightarrow NN^*(1535)}|^2.
\end{equation}
This introduction of isospin asymmetry in the production of $N^*(1535)$ and consequently of $\eta$ mesons greatly improves agreement with experimental data from \cite{Calen:1998vh} as seen in Fig.~\ref{fig:eta_prod_top}.

\subsection{Dilepton production} \label{sec:dil_prod}

Dileptons in SMASH are solely produced by direct or Dalitz decays of resonances (Table~\ref{decay_table}).

\begin{table}[h]
\caption{Direct and Dalitz dielectron decays}
\begin{center}
\begin{tabular}{c}
\hline
$\rho\rightarrow e^+e^-$\\
$\omega\rightarrow e^+e^-$\\
$\phi\rightarrow e^+e^-$\\
\hline
$\pi\rightarrow e^+e^- \gamma$\\
$\eta\rightarrow e^+e^- \gamma$\\
$\eta'\rightarrow e^+e^- \gamma$\\
$\omega\rightarrow e^+e^- \pi^0$\\
$\phi\rightarrow e^+e^- \pi^0$\\
$\Delta^+\rightarrow e^+e^- p$\\
$\Delta^0\rightarrow e^+e^- n^0$\\
\hline
\end{tabular}
\end{center}
\label{decay_table}
\end{table}%
The vector mesons $\rho$, $\omega$ and $\phi$ decay directly into a lepton pair, therefore the invariant mass of the pair equals the mass of the resonance. Although in principle direct decays of either electrons and muons are implemented, this work will focus on dielectrons only. Results of the dimuon production can be found in \cite{staude:2015bs}. Dalitz decays are incorporated for pseudo scalar mesons ($\pi$,$\eta$,$\eta'$), vector mesons ($\omega$, $\phi$), and $\Delta$ baryons. All resonances are either produced by inelastic scattering, 2$\rightarrow$1 absorption, or decays of other resonances. This also means that directly decaying resonances, like, e.g., the $\rho$ meson, include Dalitz-like contribution by coupling to baryonic resonances via processes like $N^*/\Delta^*\rightarrow \rho X\rightarrow e^+e^-X$ (see section~\ref{sec:pp}). Note that in addition to the modifications of the vector meson spectral functions due to the described explicit coupling to baryonic degrees of freedom, collisional broadening is dynamically taken into account by construction. In a dense hadronic medium, the chance for the resonances with electromagnetic decay channels to scatter with another particle before they decay is enhanced compared to reactions in the vacuum. This leads to a reduction of the lifetime and therefore an effective broadening. All dilepton decays are treated isotropically. Other transport models \cite{Bratkovskaya:2013vx, Weil:2012ji} describing dilepton production for low energies additionally include non-resonant production of dileptons via NN and $\pi$N Bremsstrahlung. Such contributions are neglected in this work, but remain a possible extension of this approach in the future.

\subsubsection{Decay widths and form factors for dilepton decays \label{sec:dw_dil}}

The decay width for direct decays $\Gamma_{V\rightarrow l^+l^-}(m_{ll})$ with $V=\rho,\omega,\phi$ under the assumptions of \emph{Vector Meson Dominance} \cite{Li:1996mi} is

\begin{align}
	\Gamma_{V \rightarrow l^+ l^-}(m_{ll}) = \frac{\Gamma_{V \rightarrow l^+ l^-}(M_0)}{M_0} \frac{M_0^4}{m_{ll}^3}\nonumber\\
	\times\sqrt{1-\frac{4m_l^2}{m_{ll}^2}}\left( 1+\frac{2m_l^2}{m_{ll}^2} \right)
\end{align}
with $m_{ll}$ being the invariant mass of the lepton pair, $M_0$ the pole mass of the vector meson and $m_l$ the lepton mass. For $\Gamma_{V \rightarrow l^+ l^-}(M_0)$ values from the PDG \cite{Agashe:2014kda} are used. The $m_{ll}^{-3}$ dependence can also be observed in Fig.~\ref{fig:spectral} for contributions to the spectral function below the hadronic threshold (red dashed line), where only the dilepton width is contributing.

For Dalitz decays the invariant mass of the dilepton is not fixed by the mass of the decaying resonance, because of the three decay products. Hence only a differential decay width $d\Gamma/dm_{ll}$ is given. For the pseudoscalar Dalitz decays $P = \pi^0,\eta,\eta'$ the differential width is given by \cite{Landsberg:1986fd},

\begin{equation}
	\frac{d\Gamma_{P\rightarrow\gamma e^+e^-}}{dm_{ll}} = \frac{4\alpha}{3\pi}\,\frac{\Gamma_{P\rightarrow\gamma\gamma}}{m_{ll}} \left( 1 - \frac{m_{ll}^2}{m_P^2} \right)^3 |F_P(m_{ll})|^2
\end{equation}
with $\Gamma_{\pi^0\rightarrow\gamma\gamma} = 7.6\cdot 10^{-6}$ MeV, $\Gamma_{\eta\rightarrow\gamma\gamma} = 5.2\cdot 10^{-4}$ MeV and $\Gamma_{\eta'\rightarrow\gamma\gamma} = 4.4\cdot 10^{-3}$ MeV \cite{Agashe:2014kda}, $\alpha=1/137$ and $m_P$  the mass of the pseudosalar meson. The form factors $F_P$ are

\begin{equation}
	F_{\pi^0}(m_{ll}) = 1 + b_{\pi^0}m_{ll}^2\,,\,\,\,\,\,\,\,b_{\pi^0} = 5.5\,\textrm{GeV}^{-2}\,,
\end{equation}

\begin{equation}
	F_{\eta}(m_{ll}) = \left(1-\frac{m_{ll}^2}{\Lambda_\eta^2}\right)^{-1}\,,\,\,\,\,\,\,\,\Lambda_\eta = 0.716\,\textrm{GeV}
\end{equation}
with $\Lambda_\eta$ taken from \cite{Arnaldi:2009aa}. For the $\eta'$ form factor the QED approximation $F_{\eta'}(m_{ll}) = 1$ is used. The vector-meson Dalitz decays ($V=\omega,\phi$) are parametrized by \cite{Landsberg:1986fd, Bratkovskaya:1996qe},

\begin{align}
  \label{eq:omega_dalitz}
	\frac{d\Gamma_{V\rightarrow\pi^0 e^+e^-}}{dm_{ll}} = \frac{2\alpha}{3\pi}\frac{\Gamma_{V\rightarrow\pi^0\gamma}}{m_{ll}}\nonumber\\
	\times\,\left[ \left(1+\frac{m_{ll}^2}{m_V^2-m_\pi^2}\right)^2 - \frac{4m_V^2m_{ll}^2}{(m_V^2-m_\pi^2)^2} \right]^{3/2} \nonumber\\
	\times\,|F_V(m_{ll})|^2\,,
\end{align}
where $m_V$ is the mass of the vector meson, $m_\pi$ the pion mass and

\begin{equation}
	|F_\omega(m_{ll})|^2 = \frac{\Lambda_\omega^4}{(\Lambda_\omega^2-m_{ll}^2)^2+\Lambda_\omega^2\Gamma_\omega^2}\,.
\end{equation}
The other parameters are set as follows: $\Gamma_{\omega\rightarrow\pi^0\gamma} = 0.703$ MeV, $\Gamma_{\phi\rightarrow\pi^0\gamma} = 5.4$ keV \cite{Agashe:2014kda}, $\Lambda_\omega = 0.65$ GeV and $\Gamma_\omega = 75$ MeV \cite{Bratkovskaya:1996qe}. For the $\phi$ form factor the QED approximation $|F_\phi(m_{ll})|^2 = 1$ is chosen. Note, that in previous work \cite{Weil:2016fxr} the possibility of describing these decays, similar as the hadronic $V\rightarrow 3\pi$ decays, in two steps via an intermediate $\rho$ meson $V\rightarrow \pi\rho\rightarrow\pi e^+e^-$ has been explored. This would render the parametrizations given above obsolete and remains an appealing option for the future. For this work we have chosen to use the more established direct treatment, which also allows more direct comparison to similar approaches \cite{Weil:2012ji, Schmidt:2008hm} that rely on the same formalism.

For the $\Delta$ Dalitz decay, the differential decay width by Krivoruchenko et al. \cite{Krivoruchenko:2001hs} is applied,

\begin{equation}
	\frac{d\Gamma_{\Delta\rightarrow N e^+e^-}}{dm_{ll}} = \frac{2\alpha}{3\pi}\frac{\Gamma_{\Delta\rightarrow N\gamma^*}(m_{ll})}{m_{ll}}\,,
\end{equation}

\begin{align}
	\Gamma_{\Delta\rightarrow N\gamma^*}(m_{ll}) = \frac{\alpha}{16}\frac{(m_\Delta+m_N)^2}{m_\Delta^3m_N^2}\nonumber \\\times\,[(m_\Delta+m_N)^2-m_{ll}^2]^{1/2}\nonumber \\
	\times\,[(m_\Delta-m_N)^2-m_{ll}^2]^{3/2}\nonumber\\
	\times\,|F_\Delta(m_{ll})|^2\,.
\end{align}
The form factor $|F_\Delta(m_{ll})|^2$ is a topic of ongoing debate~\cite{Ramalho:2015qna, Adamczewski-Musch:2017hmp}. For this work, it is chosen to be constant and fixed at the photon point $F_\Delta(0) = 3.12 \equiv F_\Delta(m_{ll})$, where it is known that $\Gamma_{\Delta\rightarrow N\gamma}(0)=702$ MeV \cite{Agashe:2014kda}.

\subsubsection{Shining method}

In experiment as well as in numerical simulations a major challenge of electromagnetic probes is their rare production. For dileptons the decay branching ratios are small, typically on the order of $10^{-5}$. Leptons are therefore treated perturbatively by the so called \emph{Time Integration Method}, also referred to as the \emph{Shining Method} \cite{Heinz:1991fn, Li:1994cj}. The idea is to obtain the dilepton yield $\Delta N_{l^+l^-}$ by integrating the decay probability of the dilepton decay mode $\Gamma_{l^+l^-} dt$ over the Lorentz-corrected ($\gamma$) lifetime $\tau=t^\dagger-t^*$ of a given resonance:

\begin{equation} \label{eq:sh}
	\Delta N_{l^+l^-} = \int_{t^*}^{t^\dagger} \frac{dt}{\gamma} \Gamma_{l^+l^-}\,.
\end{equation}

This is done numerically by continuously emitting (\emph{shining}) dileptons during the propagation of a resonance and weighting them by taking their decay probability into account. Such a perturbative treatment neglects any secondary interactions of the leptons after the decay, e.g. by the Coulomb force. This is justified since leptons only interact via the weak electromagnetic interaction and are not perturbed by strong interactions.

\section{Results} \label{sec:res}

This section is separated into four main parts covering all available experimental data for dilepton production in low beam energy collisions. First, elementary reactions are studied to confirm the hadronic baseline of the here presented transport approach. Afterwards results for nuclear systems, sorted by system size starting with pA and ending with AuAu collisions, are presented. The last part is focused on the coarse-grained hadronic space-time evolution including in-medium modifications of spectral functions. In general, contributions from all decay channels (Table~\ref{decay_table}) are taken into account, but for the Dalitz decays of $\phi$ and $\eta'$ in particular only negligible contributions are observed for the systems studied here. Therefore, almost all results exclude both channels. All results that include experimental data from HADES are filtered using the HADES acceptance filter (HAFT \cite{Janus:HAFT}). Additionally, in order to match the experimental analysis procedure an opening angle ($\Theta > 9^{\circ}$) cut and the single lepton momentum cut for the specific system have been applied.

\subsection{Elementary collisions} \label{sec:res_elem}

Elementary collisions offer the possibility to constrain and test the description of the binary reactions occurring in a nucleus-nucleus collision. Therefore, they represent a baseline for the dilepton production in nucleus-nucleus reactions. Results for three different systems are shown: proton-proton, neutron-proton and $\rm{\pi}$-proton. For the sake of data comparison, the same energies as measured by HADES were chosen~\cite{HADES:2011ab, Agakishiev:2009yf, Agakishiev:2012tc}. Results for the pion beam are not published yet, so predictions, which are not filtered for acceptance, are shown.

\subsubsection{pp \label{sec:pp}}

\begin{figure}
\includegraphics[width=0.95\columnwidth]{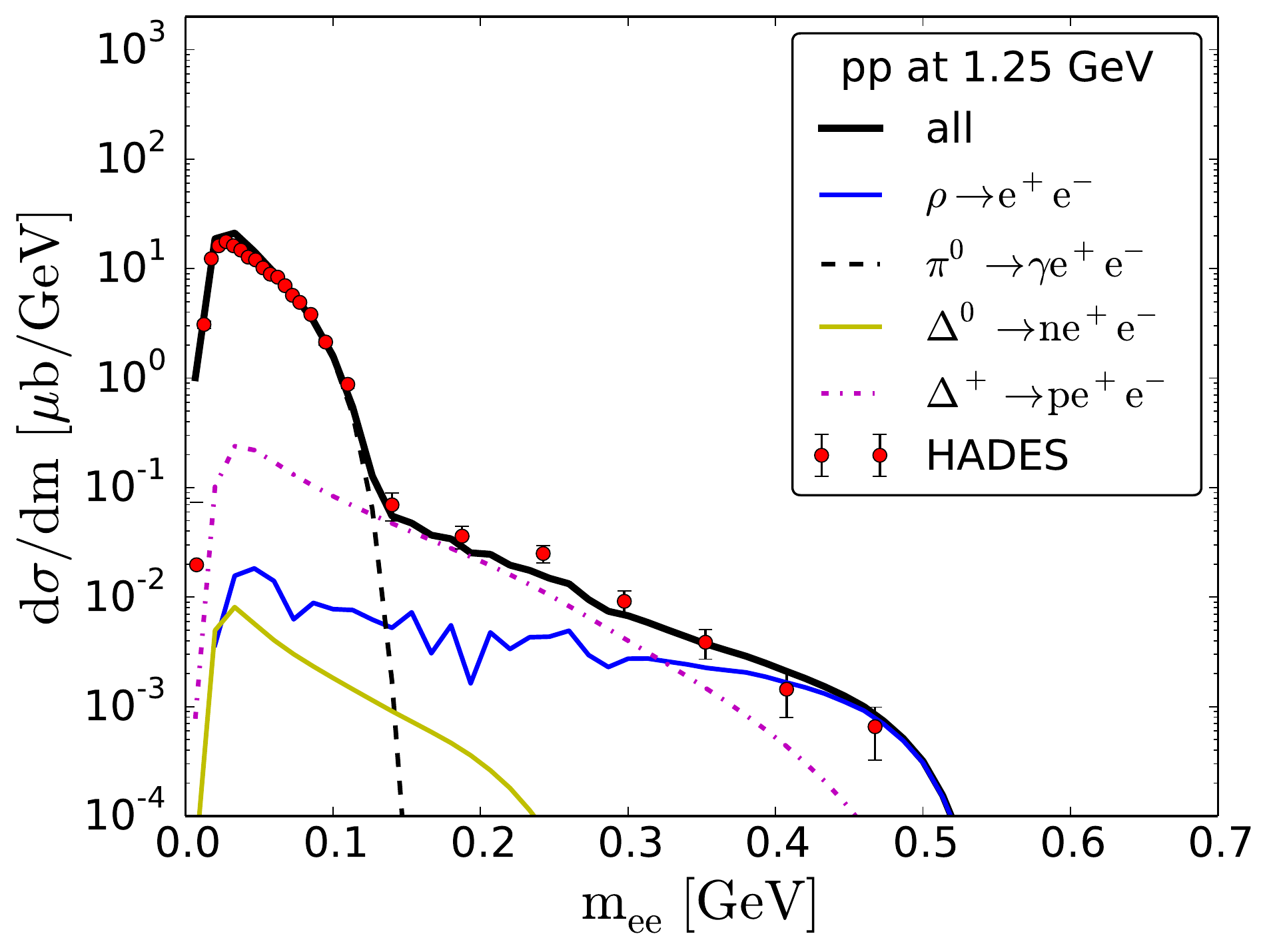}
\caption{\label{fig:pp1.22mass_top}Invariant mass spectrum of dielectrons produced in pp collisions at $E_{\rm{Kin}}=1.25\,\textrm{GeV}$. Experimental data from \cite{Agakishiev:2009yf}.}

\includegraphics[width=0.95\columnwidth]{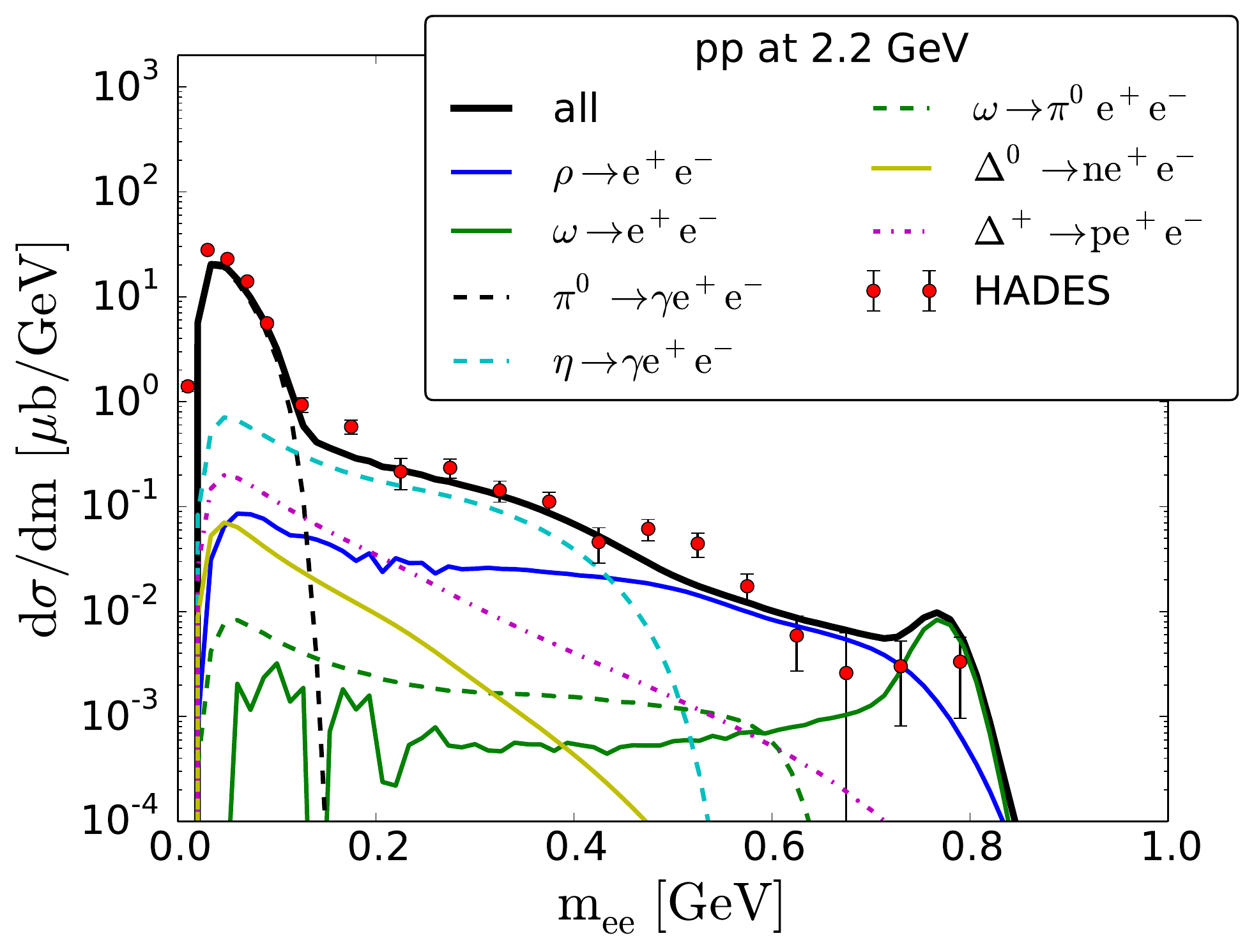}
\caption{\label{fig:pp1.22mass_bot}Invariant mass spectrum of dielectrons produced in pp collisions at $E_{\rm{Kin}}=2.2\,\textrm{GeV}$. Experimental data from \cite{Agakishiev:2012tc}.}
\end{figure}

Dilepton production in proton-proton (pp) reactions is calculated for three different kinetic energies $E_{\rm{Kin}}=1.25/2.2/3.5$ GeV in a fixed-target setup. Fig.~\ref{fig:pp1.22mass_top} and Fig.~\ref{fig:pp1.22mass_bot} show the dilepton invariant mass spectrum for the two lower energies in comparison to HADES data~\cite{Agakishiev:2009yf, Agakishiev:2012tc}.

For the reaction at $E_{\rm{Kin}}=1.25$ GeV (Fig.~\ref{fig:pp1.22mass_top}) only four different channels of the whole dilepton cocktail are contributing. The $\pi^0$ Dalitz decay dominates in the $\pi^0$ invariant mass region up to around $0.15$ GeV. Above $0.15$ GeV in the low mass region, the Dalitz decay of the $\Delta^+$ decay is dominant. Since the $\Delta^+$ and the $\Delta^0$ contributions are plotted separately a difference of more than one order of magnitude can be observed. The $\Delta^+$ is more likely to be produced, since it can be a product of the primary collision, whereas the $\Delta^0$ can only be formed in secondary reactions due to charge conservation and the fact that only $2\rightarrow2$ reactions are allowed in SMASH. In the higher invariant mass region a large contribution from the direct $\rho$ meson channels is noticed. The total yield is in good agreement with experimental data.

Since a kinetic energy of $E_{\rm{Kin}}=1.25$ GeV is slightly below and $E_{\rm{Kin}}=2.2$ GeV is above the $\eta$ production threshold, a contribution from the $\eta$ meson is seen in Fig.~\ref{fig:pp1.22mass_bot}. Also, additional significant contributions from $\omega$ decays are observed for the higher kinetic energy. Here, the $\eta$ yield is dominant for the invariant mass region up to around $0.4$~GeV. The $\pi$ again dominates for low invariant masses and the $\rho$ in the mass region above $0.4$~GeV. The peak in the $\omega\rightarrow e^+e^-$ spectrum at the $\omega$ pole mass can already be observed for this kinetic energy. The overall agreement with data is again reasonable.

\begin{figure}
\includegraphics[width=0.95\columnwidth]{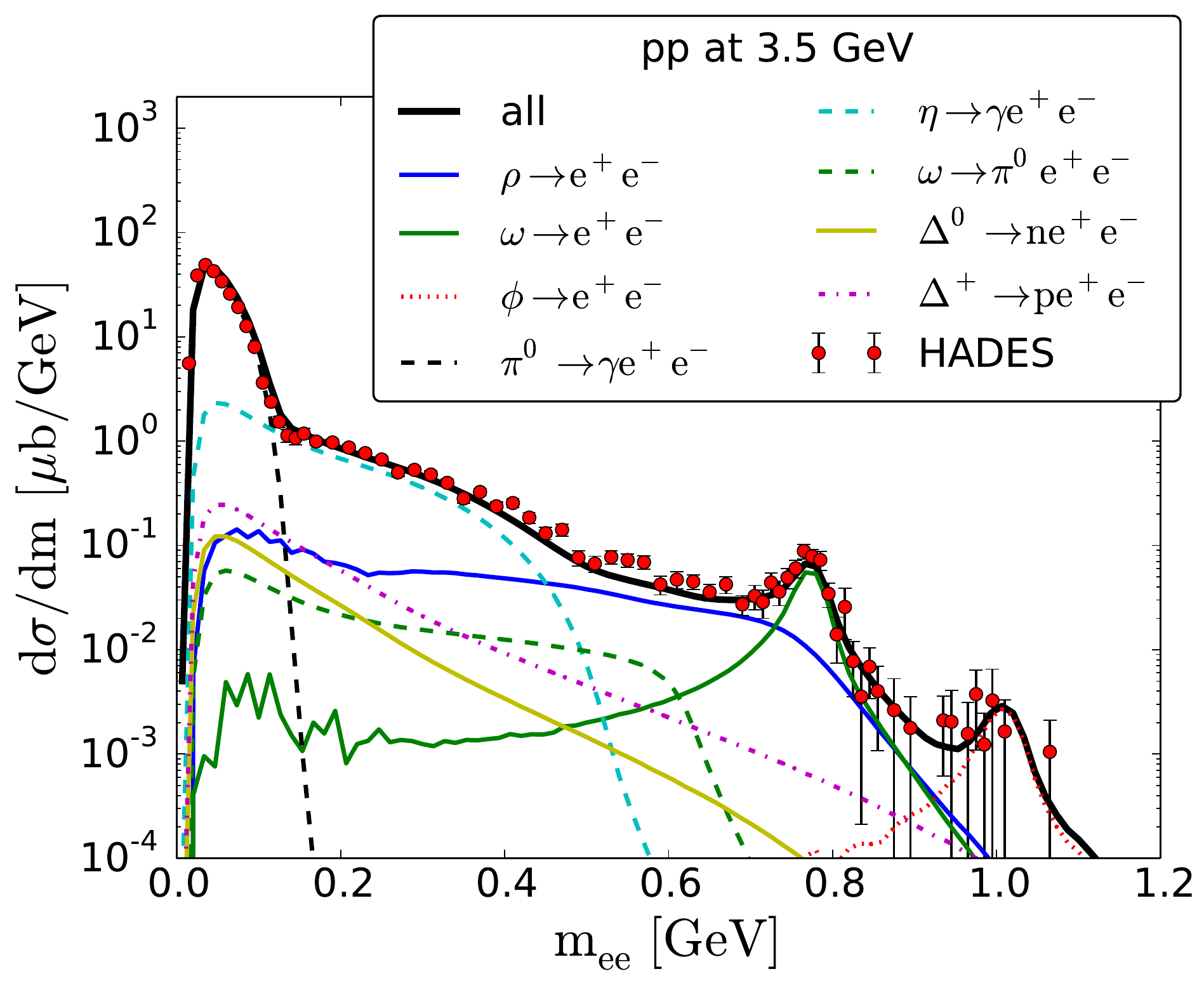}
\caption{\label{fig:pp3.5mass} Invariant mass spectrum of dielectrons produced by pp collisions at $E_{\rm{Kin}}=3.5\,\textrm{GeV}$. Experimental data from \cite{HADES:2011ab}.}
\end{figure}

Fig.~\ref{fig:pp3.5mass} shows the invariant mass spectrum produced by fixed target pp reactions with a kinetic energy of $3.5$~GeV. Because of the higher energy the spectrum reveals two new features: a more pronounced $\omega$ peak and an additional $\phi$ contribution. The significant contributions to the spectrum now reach up to $1.1$~GeV. Again, a good agreement with experimental data is observed. In Appendix~\ref{sec:apend_pp3.5}, $p_T$ and $y$ spectra for different invariant mass windows are shown for completeness.

The contributions from the direct decays of the vector mesons with masses reaching below the hadronic thresholds are important for all three energies. They originate from the resonance description introduced in Sec.~\ref{res_desc} that considers the dilepton decays for the spectral function. In the case of the $\rho$ meson, those contributions are significant for the total yield in the low mass region. For the $\omega$ they are negligible compared to e.g. the $\omega$ Dalitz decay for the here discussed elementary system. Nevertheless, Fig.~\ref{fig:pp3.5mass} does show that contributions below the hadronic threshold are observed for both mesons. This also holds for the direct $\phi$ decay, but low invariant mass contributions in pp are too small to be visible for the $\phi$ meson on the chosen scale.

After discussing the general features of dilepton production in pp collisions, the focus in the following will be on the contribution from the vector mesons.

The previously shown findings for the elementary production cross-section of $\rho$ mesons measured in the hadronic channel (Fig.~\ref{fig:xs_rho}) translate directly to the corresponding dilepton spectra. At low energies ($E_{\rm{Kin}}=1.25$ and $2.2$ GeV), where the overestimated exclusive $\rho$ production via $pp\rightarrow pp\rho$ is dominant, the $\rho$ yield is large, meaning it pushes the total yield to the upper limits of the error bars, whereas for the highest kinetic energy ($E_{\rm{Kin}}=3.5$ GeV) this is not the case. Overall, the overestimation of the exclusive cross-section does not result in an overestimation of the total dilepton yield, which indicates that the inclusive $\rho$ production is in line with the experimental dilepton results.

\begin{figure}
\includegraphics[width=0.95\columnwidth]{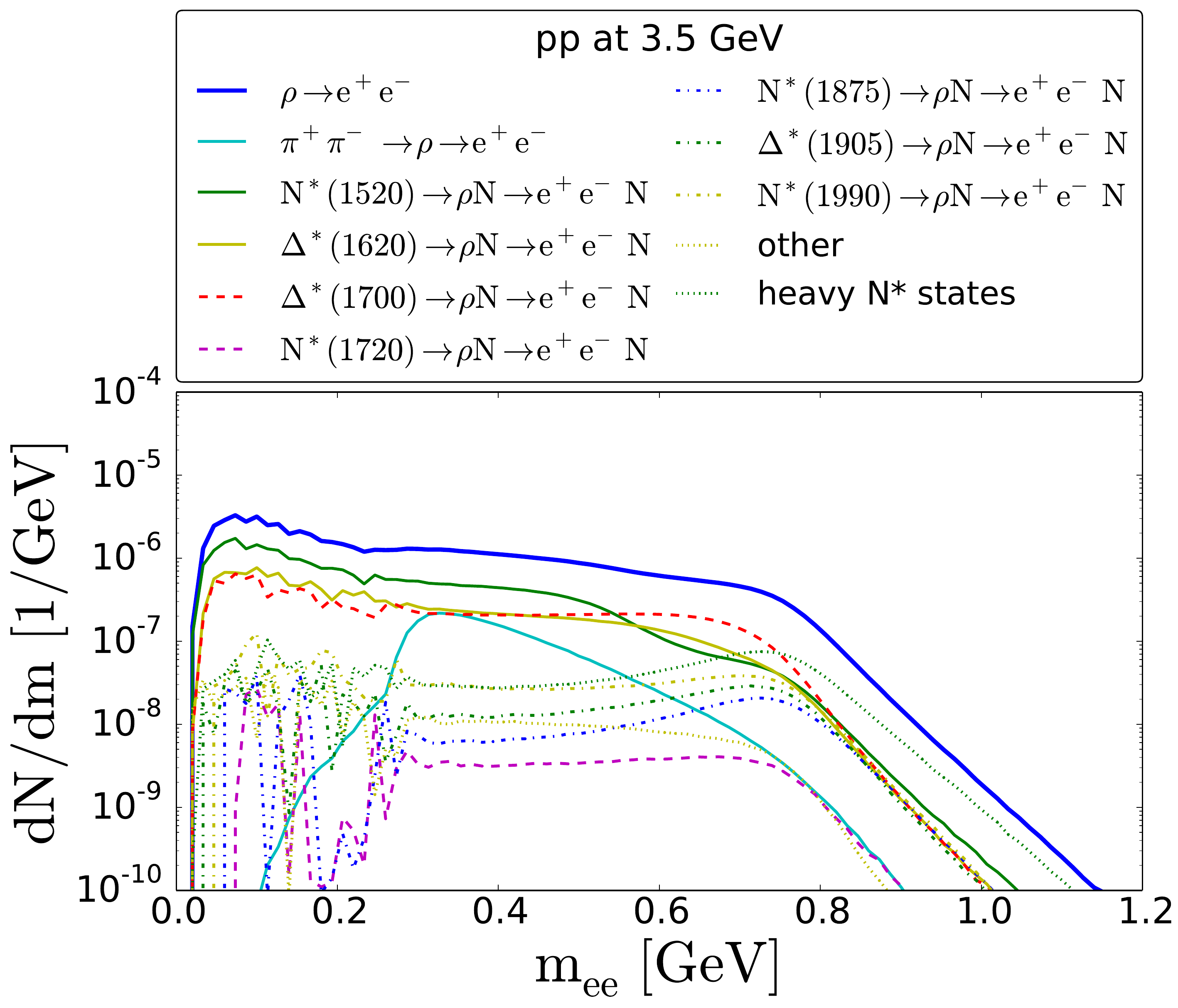}
\includegraphics[width=0.95\columnwidth]{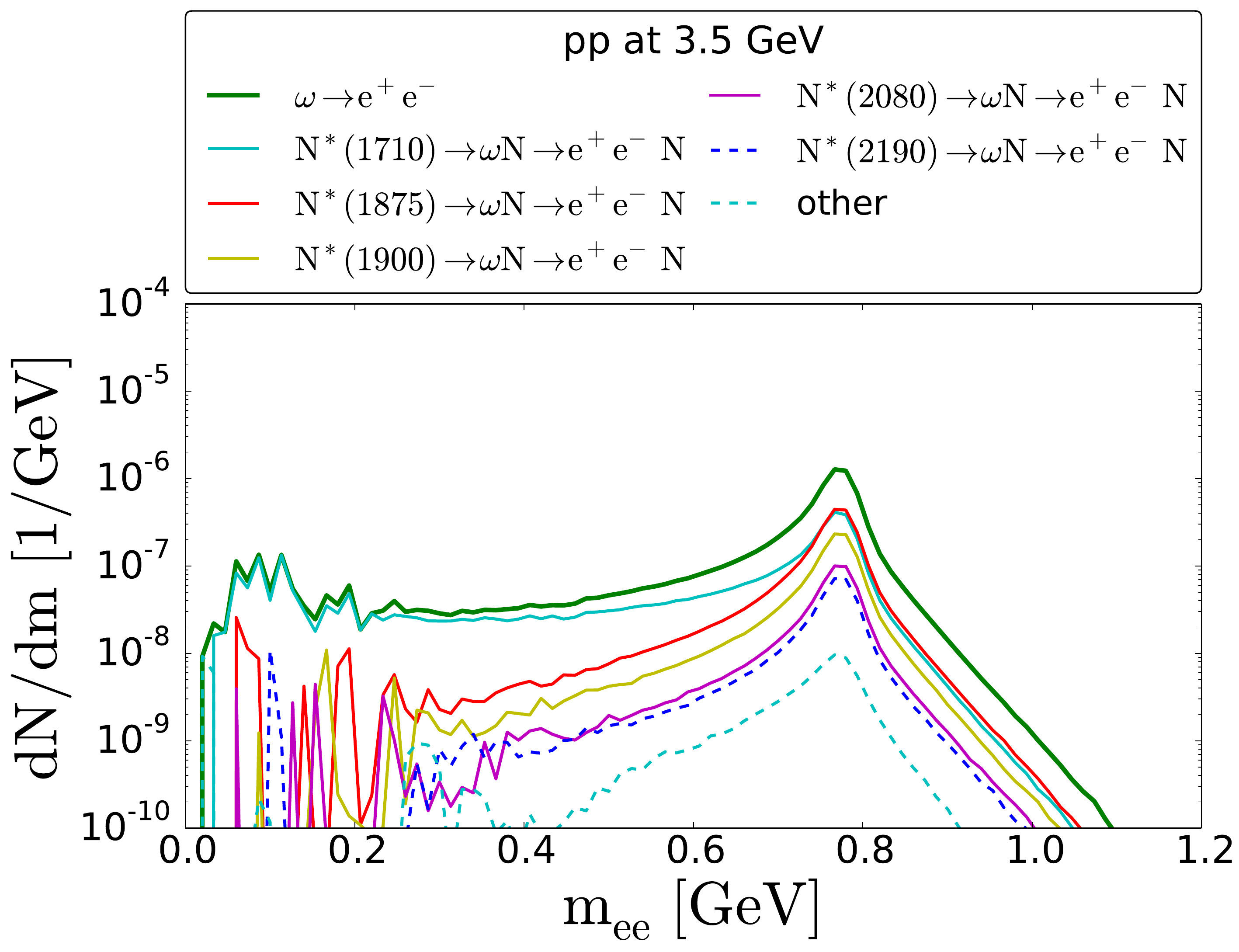}
\caption{\label{fig:pp3.3origin}Different contributions to the invariant mass spectrum of dielectrons produced by decays of $\rho$ (top) and $\omega$ (bottom) mesons for pp collisions at $E_{\rm{Kin}}=3.5\,\textrm{GeV}$. Heavy N* states include contributions from $N^*(2080),\,N^*(2190),\,N^*(2220),\, N^*(2250)$, all of the form $N^*\rightarrow\rho N\rightarrow e^+e^-N$.}
\end{figure}

In order to fully understand the dilepton production, the origin of the $\rho$ and $\omega$ resonance is investigated. Earlier studies~\cite{Weil:2013mya,Weil:2012ji,Rapp:1999us, Rapp:2000pe} revealed that in particular the coupling of the vector mesons to baryonic resonances is of importance. Such information is however challenging to obtain in experiment alone, therefore only few experimental studies are available~\cite{Agakishiev:2014wqa}. Comparisons to theoretical models that keep track of the whole process history enable insights into important couplings by splitting the $\rho$ and $\omega$ contributions by origin. Additionally, such studies offer the possibility to constrain resonance properties such as branching ratios.

At the top of Fig.~\ref{fig:pp3.3origin} the different contribution to the overall $\rho$ dilepton yield (thick blue (upper) line, same as in Fig.~\ref{fig:pp3.5mass}) are shown for pp reactions at a kinetic energy of $3.5$ GeV. To allow comparisons to the overall invariant mass spectra, all dileptons in Fig.~\ref{fig:pp3.3origin} are also acceptance filtered. Two different processes are important: $\pi^+\pi^-$ annihilation and the decays of different baryonic resonances. The annihilation process has a small yield as expected in elementary pp collisions, because it requires rare secondary scatterings. While the $\pi\pi$  process of course has a threshold at $2m_{\pi}$, the significant contributions below this threshold come from Dalitz-like contribution of the lighter baryonic resonances ($B^*\rightarrow\rho N\rightarrow e^+e^-N$, $B = N,\Delta$), mainly $N^*(1520), \Delta^*(1620)$ and  $\Delta^*(1700)$. These populate the large low-mass tail of the overall $\rho$ yield.  The different shape of the $N^*(1520)$ is due to the kinematic limitation that the pole mass of the resonance is too small to produce a $\rho$ at the pole mass in the reaction $N^*\rightarrow N\rho$. For higher invariant masses, higher baryonic resonances are important.  Especially, the combined heavy $N^*$ states ($N^*(2080),\,N^*(2190),\,N^*(2220),\, N^*(2250)$) dominate the high mass tail. Other contributions include higher mesonic states and baryonic resonances that have no significant effect on the overall $\rho$ contribution.

Fig.~\ref{fig:pp3.3origin} also shows the different contributions to the overall $\omega$ yield (thick green (upper) line, same as in Fig.~\ref{fig:pp3.5mass}) at the bottom. Since the $\omega$ width is much smaller, it shows a very clear peak structure at its pole mass in the invariant mass spectrum; $\omega$s are mainly produced by nucleon resonance decays. A clear mass ordering can be observed. The lightest baryonic resonances, $N^*(1710), N^*(1875)$, have the largest contributions followed by the heavier resonances in order of their pole masses. The contribution below the hadronic thresholds that mainly forms the  low-mass tail is the $N^*(1710)$ resonance, which is also the lightest resonance that can decay into $\omega$ with a pole mass below the $m_N+m_{\omega}$ threshold.

Overall the dilepton production in SMASH for pp collisions as the cleanest probe for elementary collisions is well understood and in good agreement with experimental data and offers a solid base to study larger systems.

\subsubsection{np\label{sec:np}}

The next system of interest is the elementary neutron-proton (np) system. Dilepton production has been measured by HADES, realizing np collisions by using a deuteron beam on a proton target and triggering on forward-going protons (so called spectator protons). These reactions are called \emph{quasi-free}. Since the deuteron is a bound system, the nucleons inside carry additional momentum. The results for this work were obtained following an ansatz referenced in the HADES publication \cite{Agakishiev:2009yf}. The neutron projectile is given additional momentum according to the momentum distribution of the PARIS potential \cite{AbdelBary:2006zz, Lacombe:1981eg}, neglecting the relatively small binding energy of the deuteron itself.

\begin{figure}
    \includegraphics[width=0.95\columnwidth]{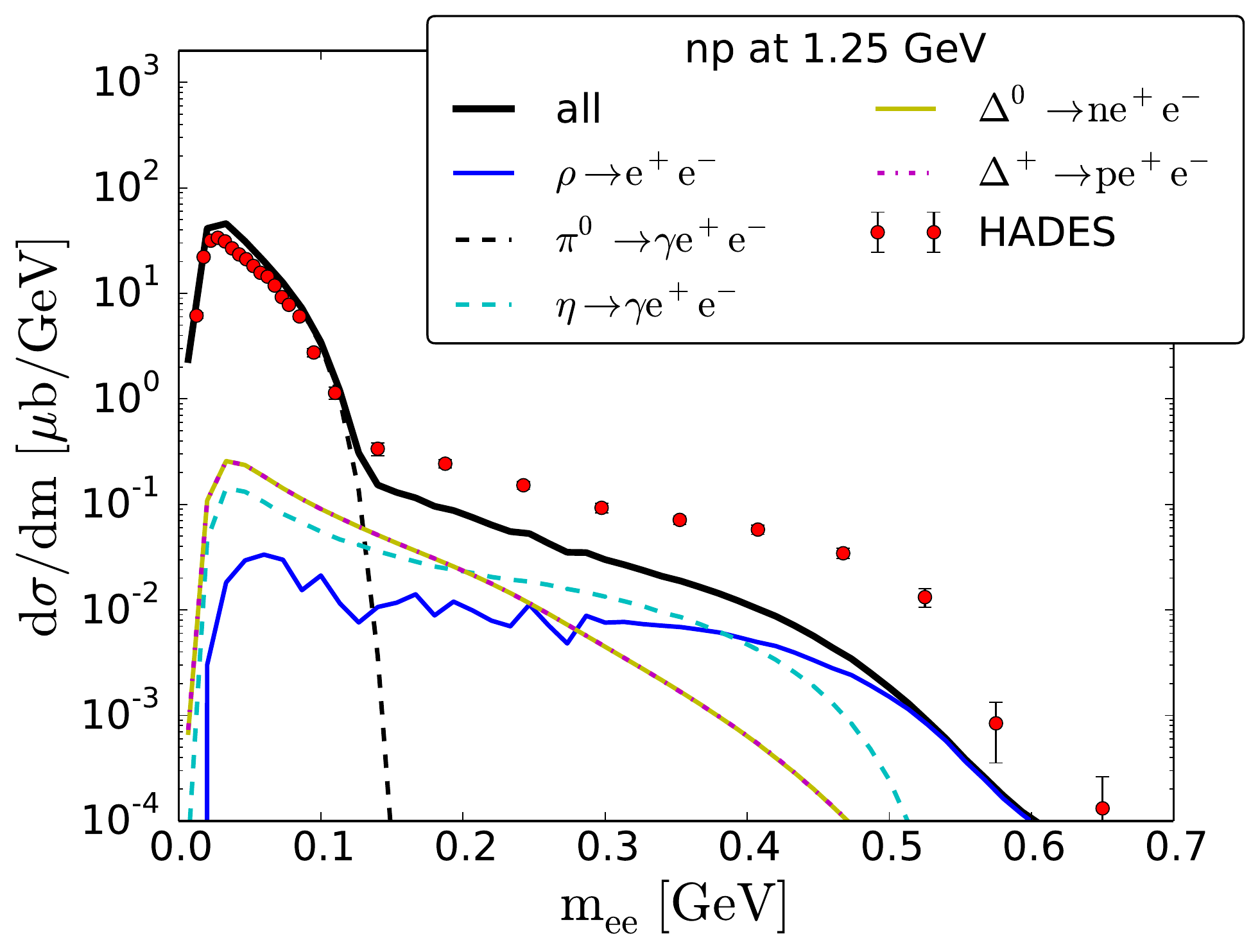}
    \caption{\label{fig:npmass} Invariant mass spectrum of dielectrons produced by quasi-free np reactions at $E_{\rm{Kin}}=1.25\,\textrm{GeV}$. Experimental data from \cite{Agakishiev:2009yf}}
\end{figure}

Fig.~\ref{fig:npmass} shows results for the dilepton production of quasi-free np reactions at $E_{\rm{Kin}}=1.25\,\textrm{GeV}$. The additional momentum of the neutron inside the deuteron leads to a higher kinematic threshold than in the pp case (Fig.~\ref{fig:pp1.22mass_top}) for the same energy, so contributions up to $0.6$~GeV are significant. The same channels as in pp are contributing, but, because the energy now reaches above the $\eta$ threshold, an additional $\eta$ yield is observable. In addition, the isospin asymmetry between $\Delta^+$ and $\Delta^0$ does not exist anymore, since both are equally likely to be excited in a primary collision. The mass region below $0.15$~GeV is dominated by the $\pi^0$ contribution. In the low-mass region, $\eta, \Delta$ and $\rho$ are all contributing, while for masses above $0.4$~GeV, the direct $\rho$ decay becomes the dominant contribution.

Compared to the HADES data \cite{Agakishiev:2009yf}, a large discrepancy is observed for masses higher than $0.15$~GeV, suggesting that the $\pi^0$ contribution is described reasonably well, but other channels are underestimated. One possible extensions that would enhance the total yield include the addition of np Bremsstrahlung or a $np\rightarrow d\eta$ channel. Such extensions, however, were not successful in describing the experimental data in the similar GiBUU transport model \cite{Weil:2012ji}. Two other promising explanation are tested in \cite{Adamczewski-Musch:2017oij} in the context of the same experimental data. The first focuses on the, in this work neglected, radiation of dileptons from an internal charged meson line and is based on a one-boson exchange model \cite{Shyam:2010vr}. The second focuses on double $\Delta$ excitations as a possible solution \cite{Bashkanov:2013dua}. Both models lead to a significant better agreement with experimental data, with the first one slightly favored by the experimental data~\cite{Adamczewski-Musch:2017oij}. Additionally, as argued in \cite{Martemyanov:2011hv} the channel $np\rightarrow d e^+e^-$ might be important for the spectrum discussed here.

Similar to other transport approaches \cite{Weil:2012ji, Bratkovskaya:2013vx}, the np system seems to be only underestimated at this  low energy. The later discussed carbon-carbon collisions for example, which are close to a superposition of pp and np collisions~\cite{Agakishiev:2009yf}, only show a similar systematic underestimation of the dilepton production around the same energy of $1A$~GeV. For the higher  discussed energy ($E_{\rm{Kin}}=2.0A\,\textrm{GeV}$), the agreement with experimental measurements improves considerably. A reasonable agreement is also seen for higher energies in the studied proton-nucleus system. Improvements as discussed above are therefore left for future work.

\subsubsection{Pion beam}

Besides the discussed NN reactions, pion-beam reactions, where  $\pi^-$ scatter on a proton target, are considered. The kinetic energy of $E_{\rm{Kin}}=0.56\,\textrm{GeV}$ matches upcoming HADES results \cite{Scozzi:2017sho} for this system and is specifically chosen to probe the $\rho$ production around the $N^*(1520)$ pole mass.

\begin{figure}
    \includegraphics[width=0.95\columnwidth]{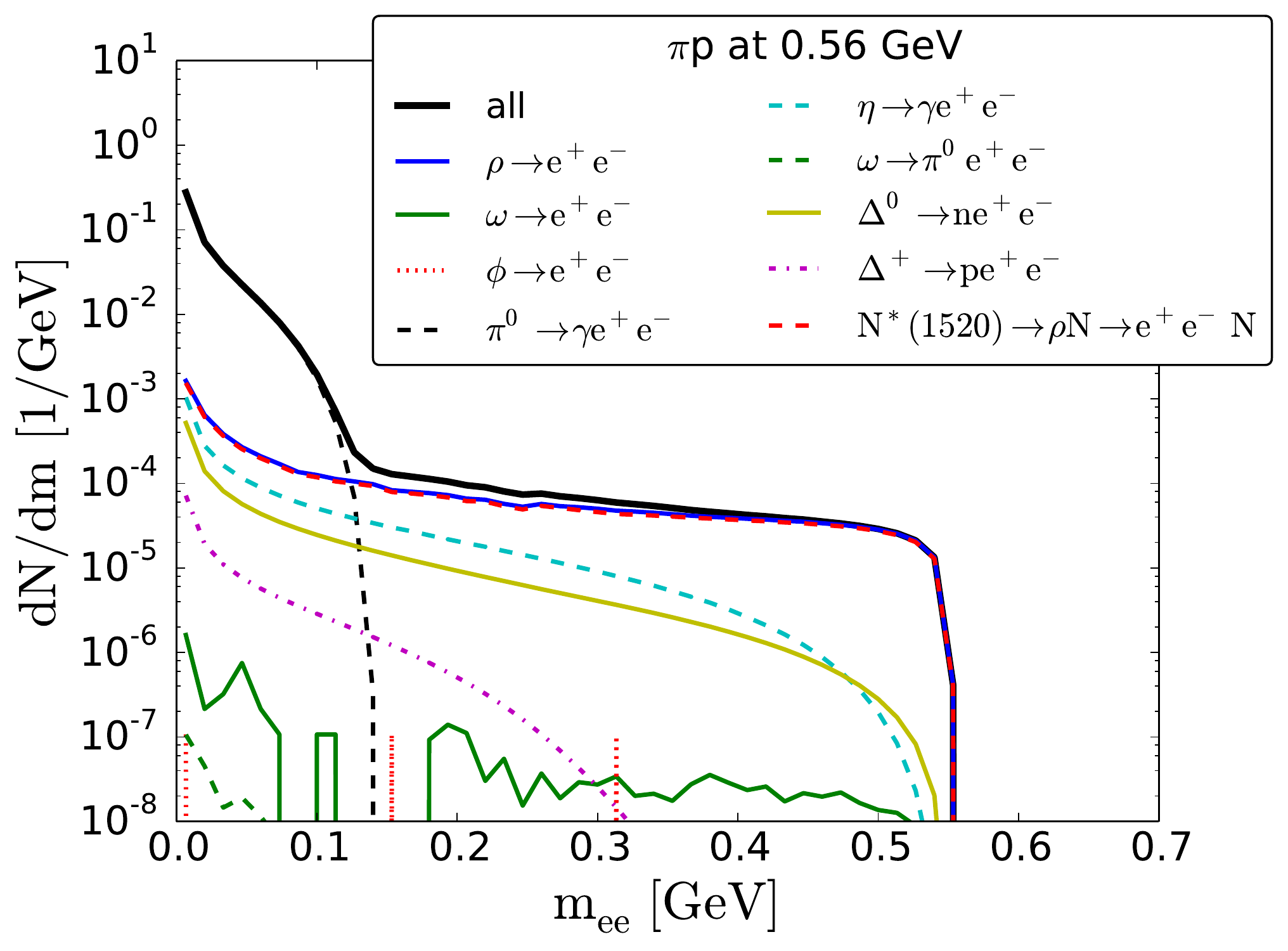}
    \caption{\label{fig:pipmass} Invariant mass spectrum of dielectrons produced by $\rm{\pi}$p reactions at $E_{\rm{Kin}}=0.56\,\textrm{GeV}$.}
\end{figure}

Indeed, the $\rho$ dilepton decay is observed to be the dominant contribution to the dilepton invariant mass spectrum from $\rm{\pi}$p at $E_{\rm{Kin}}=0.56\,\textrm{GeV}$ (Fig.~\ref{fig:pipmass}). Only for invariant masses lower than $0.15$~GeV, it is exceeded by the $\pi$ decay contribution. Other smaller contributions include $\eta$ and $\Delta^0$, negligible are $\Delta^+$, $\omega$ and $\phi$. Compared to pp the ordering of the $\Delta^+$ and $\Delta^0$ is inverted for this system due to the same reason as mentioned before: because of charge conservation only $\Delta^0$ can be produced in primary collisions. The sharp kinematic threshold at $0.56$ GeV due to the available center of mass energy is noticeable as well. The red dashed line in Fig.~\ref{fig:pipmass} indicates the $N^*(1520)$ contribution to the $\rho$ spectrum. It is the only relevant contribution to the $\rho$ spectrum and with this to the overall spectrum above $0.15$~GeV. Therefore this setup provides a good opportunity to test and constrain the coupling of the $\rho$ to the $N^*(1520)$. Currently SMASH treats the $N^*(1520)$ Dalitz decay ($N^*(1520)\rightarrow e^+e^-N$) via the strict Vector Meson Dominance assumption, where the resonance decays via an intermediate $\rho$ meson. The pion beam experimental data will be valuable to constrain this assumption and the extensive theoretical investigation of the $\rho -N$ interaction in general, which started with~\cite{Friman:1997tc,Peters:1997va,Rapp:1997ei} (see~\cite{Rapp:1999ej} for a review). For example, a test of the simple "QED point-like" $R\gamma^*$ model~\cite{Zetenyi:2001fu}, where the Dalitz decay is performed directly and the involved electromagnetic form factor is chosen to be constant, would be possible.

\subsection{Proton-nucleus collisions} \label{sec:res_pA}

The dilepton production in proton-nucleus (pA) collisions, as a cold nuclear matter scenario, is discussed for a proton projectile that scatters on a niobium target with $E_{\rm{Kin}}=3.5\,\textrm{GeV}$ (pNb). To obtain the cross-section the value reported in~\cite{Agakishiev:2012vj} ($\sigma_{\rm{pNb}}=848\pm127$ mb)  is used.

\begin{figure}
    \includegraphics[width=0.95\columnwidth]{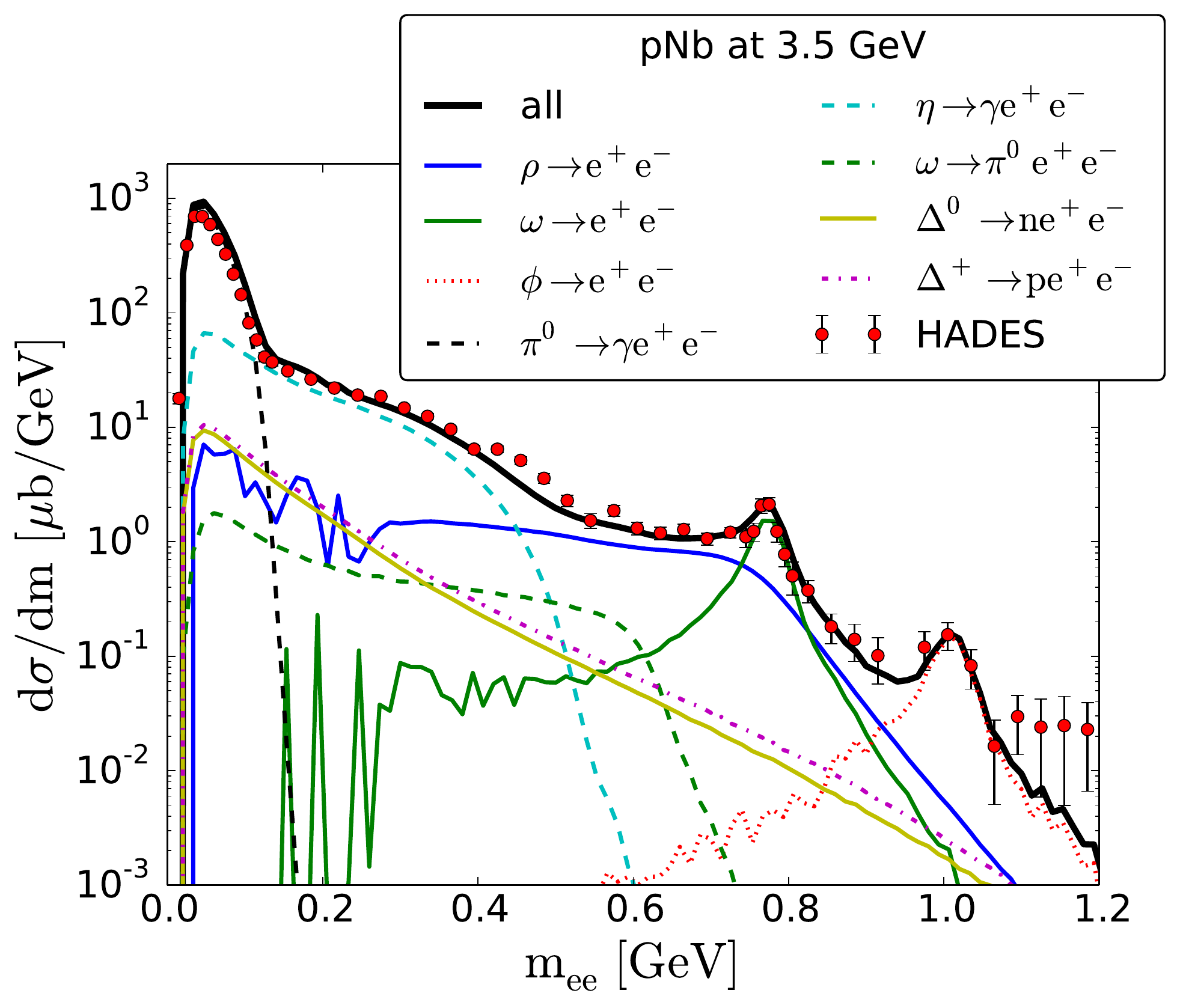}
    \caption{\label{fig:pNbmass} Invariant mass spectrum of dielectrons produced by pNb reactions at $E_{\rm{Kin}}=3.5\,\textrm{GeV}$. Experimental data from \cite{Agakishiev:2012vj}.}
\end{figure}

A comparison to HADES data \cite{Agakishiev:2012vj} of the invariant electron-pair mass spectrum is displayed in Fig.~\ref{fig:pNbmass}. In the low-mass region, the $\pi$ and $\eta$ contributions are prominent. The $\pi$ peak is slightly overestimated, hinting at a problem with the overall normalization. The experimental data also reveal a stronger shoulder around $0.5$~GeV.

Besides the dominant $\rho$ and $\omega$ contributions in the mass region from $0.5$~GeV up to $0.9$~GeV, a strong $\phi$ peak surfaces around $1$ GeV. The $\phi$ production is experimentally not well known. Neither $N^* \to \phi N$ branching ratios nor the $pp \to pp\phi$ cross section beyond the threshold are constrained by experimental data. Therefore, the simple ansatz reported in \cite{Steinheimer:2017dtk} is followed in this work. All nucleon resonances with a pole mass of $2080$ MeV or higher decay into $N\phi$ with a fixed branching ratio. The $\phi$ peak in pNb allows us to constrain the fixed branching ratio within our approach to a value of $0.5\%$, which is larger than the value reported in~\cite{Steinheimer:2017dtk}. Note that because of the larger system medium effects like absorption also play a role. Unfortunately, the corresponding elementary pp data that would offer a clean vacuum reference only provides an upper bound in the $\phi$ peak mass region due to large error bars. Nevertheless, the obtained value is consistent with this bound, since an agreement within error bars is seen in Fig.~\ref{fig:pp3.5mass}.

Overall experimental data and SMASH results are in reasonable agreement. Therefore, it seems that the resonance description based on vacuum properties is able to account for the dynamics to some extent, whereas the underestimation around $0.5$~GeV might hint at an onset of a broadening of the $\rho$-like contribution due to a stronger coupling of the $\rho$ to baryonic resonances.

\begin{figure}
    \includegraphics[width=0.95\columnwidth]{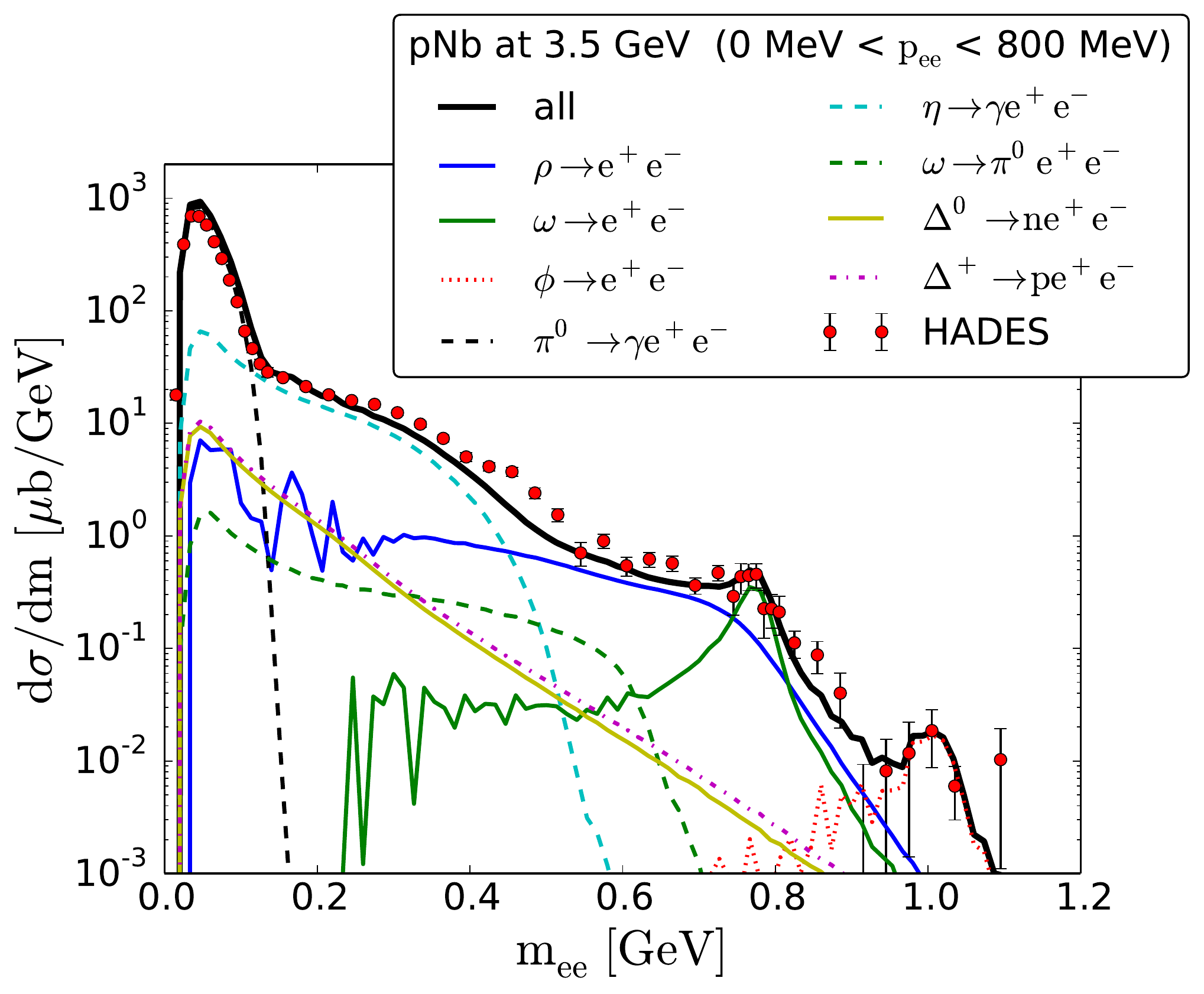}
    \includegraphics[width=0.95\columnwidth]{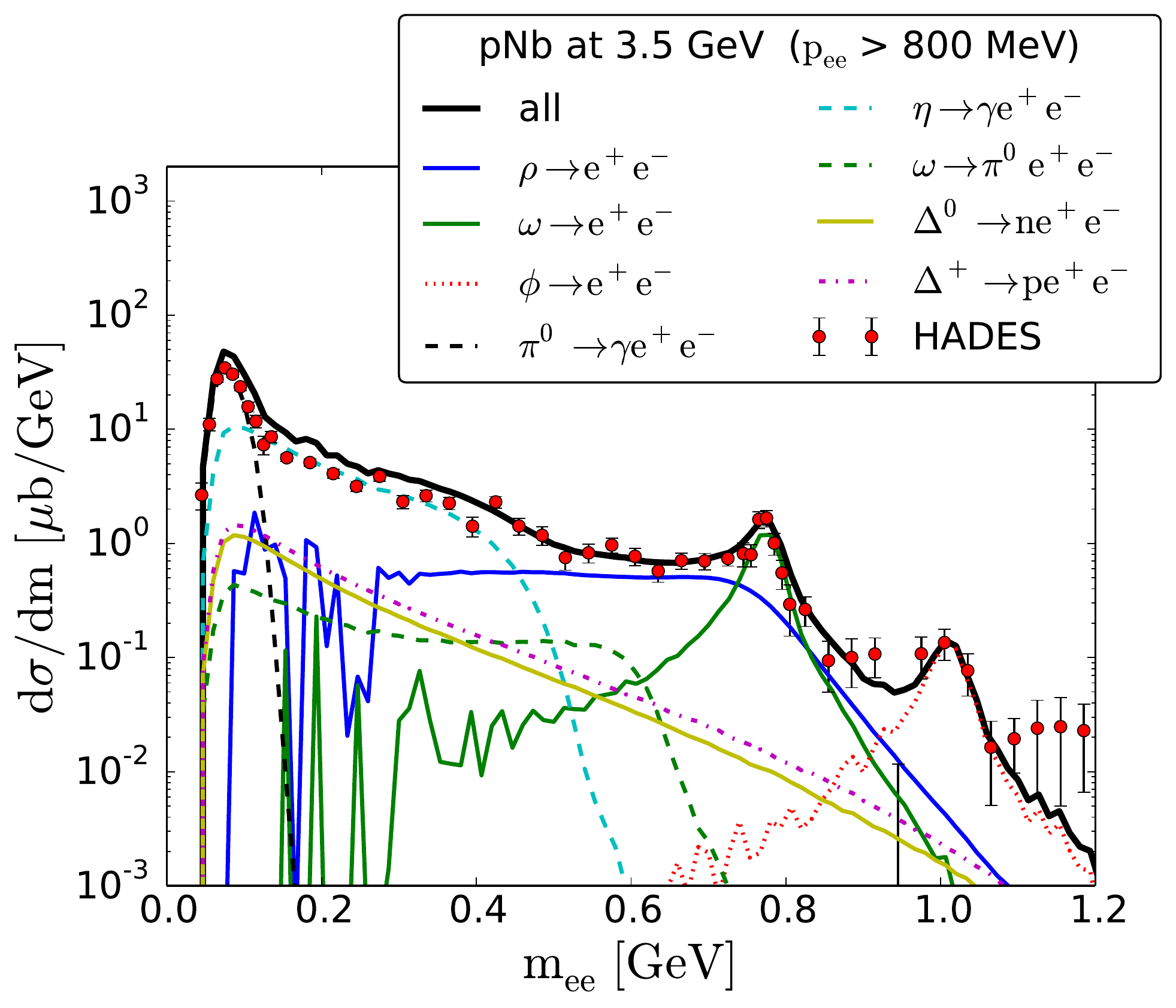}
    \caption{\label{fig:pNbdiff} Invariant mass spectrum of dielectrons produced by pNb reactions at $E_{\rm{Kin}}=3.5\,\textrm{GeV}$ in different dielectron momentum ($\rm{p_{ee}}$) windows. Experimental data from \cite{Agakishiev:2012vj}.}
\end{figure}

Fig.~\ref{fig:pNbdiff} displays the invariant mass spectrum for two different dilepton momentum windows ($0 < \rm{p_{ee}} < 800$ MeV and $\rm{p_{ee}} > 800$ MeV). Again, an overall reasonable agreement is observed. The underestimation of the shoulder at $0.5$~GeV is only seen for the low momentum dileptons, which further points to a broadening of the $\rho$ spectral shape. Effects by the medium are enhanced for the low momentum dileptons, since the decaying resonances are traversing the medium longer. The $\omega$ and the $\phi$ peak are nicely matched. On the one hand, this validates the extracted $N^* \to \phi N$ branching ratio. On the other hand, if one compares the peaks between momentum cuts, the peaks are suppressed for the low momentum dileptons. This is caused by absorption of low-momentum resonances inside the cold nuclear matter. These findings support the results reported in \cite{Agakishiev:2012vj}. Furthermore, this validates the microscopic dynamics, since absorption is a medium effect that is intrinsic to transport approaches (in contrast to modification of spectral shapes).

Furthermore, the relatively solid description of the experimental data hints that the dilepton production cross section in np collision at this energy is in better agreement with experimental results than for the lower energy of $1.25$ GeV discussed in Sec.~\ref{sec:pp}, since initial pp and np reactions are roughly equally contributing to the pA yield.

The here reported findings for the proton-induced reactions (including the results from Sec.~\ref{sec:pp} and Sec.~\ref{sec:np}) align with the results from the GiBUU approach~\cite{Weil:2012ji}. 
Differences are found in the low-mass $\omega\rightarrow e^+e^-$ contributions and the composition of the $\rho\rightarrow e^+e^-$ yield. The latter is caused by the adaptation of recent PDG branchings and, in general, a different set of resonance states. Additionally, the discussion here is extended to include the spectra for the two $\rm{p_{ee}}$ windows. 
Similar efforts as in~\cite{Weil:2012ji} to improve the agreement with the np experimental data such as Bremsstrahlung are left for future work.

\subsection{Nucleus-nucleus collisions} \label{sec:res_AA}

The main goal of research in this field is to study heavy-ion collisions. In the following, results for three different nucleus-nucleus collision systems are shown and compared to experimental data, wherever available. This section also addresses the question of which spectra are sensitive to additional medium modifications by deviating from results based on vacuum properties.It builds on the above presented dilepton production in elementary reactions and cold nuclear matter.

\subsubsection{CC \label{sec:CC}}

Light nucleus-nucleus collisions offer a good starting point for studying dilepton production under the assumption of vacuum resonance properties in larger collision systems. Results in this section include invariant mass spectra of the produced dielectrons in carbon-carbon (CC) collisions for two kinetic energies: $E_{\rm{Kin}}=1.0A\,\textrm{GeV}$ and $E_{\rm{Kin}}=2.0A\,\textrm{GeV}$.

\begin{figure}
    \includegraphics[width=0.95\columnwidth]{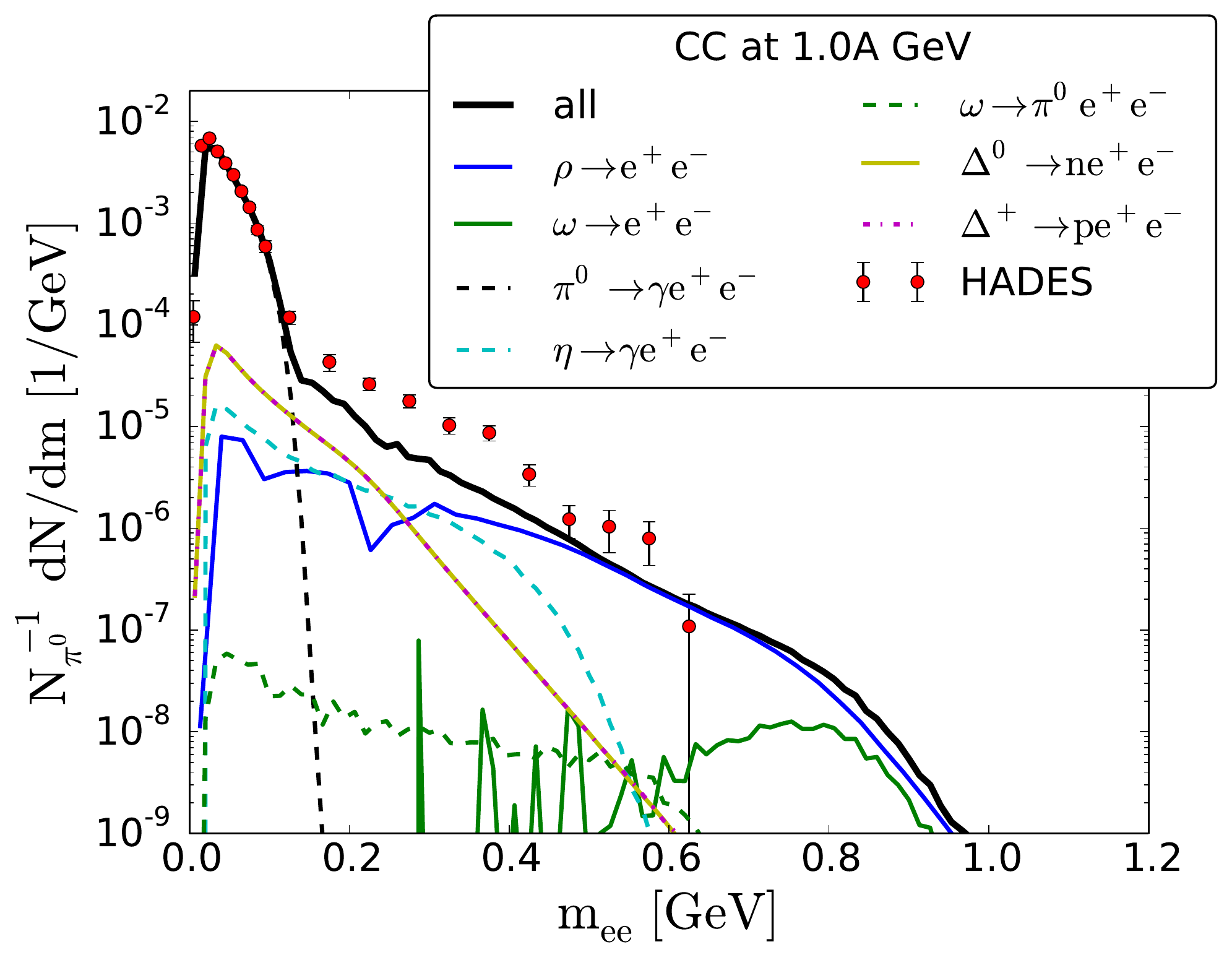}
    \caption{\label{fig:CC1mass} Invariant mass spectrum of dielectrons produced by CC reactions at $E_{\rm{Kin}}=1.0A\,\textrm{GeV}$. Data from \cite{Agakishiev:2007ts}.}
\end{figure}

Fig.~\ref{fig:CC1mass} shows the spectrum for the lower beam energy. The main contributions originate from the $\pi$, $\Delta$, $\eta$ and $\rho$ channels. The $\omega$ decays do not have a large impact on the overall yield. The $\Delta^0$ and $\Delta^+$ yields are lying on top of each other, due to the equal numbers of protons and neutrons, and therefore similar production probability. The comparison with HADES data \cite{Agakishiev:2007ts} reveals a disagreement in the low mass region between $0.15$~GeV and $0.4$~GeV. Even though the shape potentially matches the data, the total yield is underestimated. This can be understood recalling the previously discussed elementary results in Sec.~\ref{sec:pp} and Sec.~\ref{sec:np}. The dilepton production in pp collisions is in good agreement with data, but too few dileptons are produced by np collisions, in consequence an underestimation around the same kinetic energy is expected, since CC is known to be close to a mere superposition of binary NN reactions \cite{Agakishiev:2009yf}.

\begin{figure}
    \includegraphics[width=0.95\columnwidth]{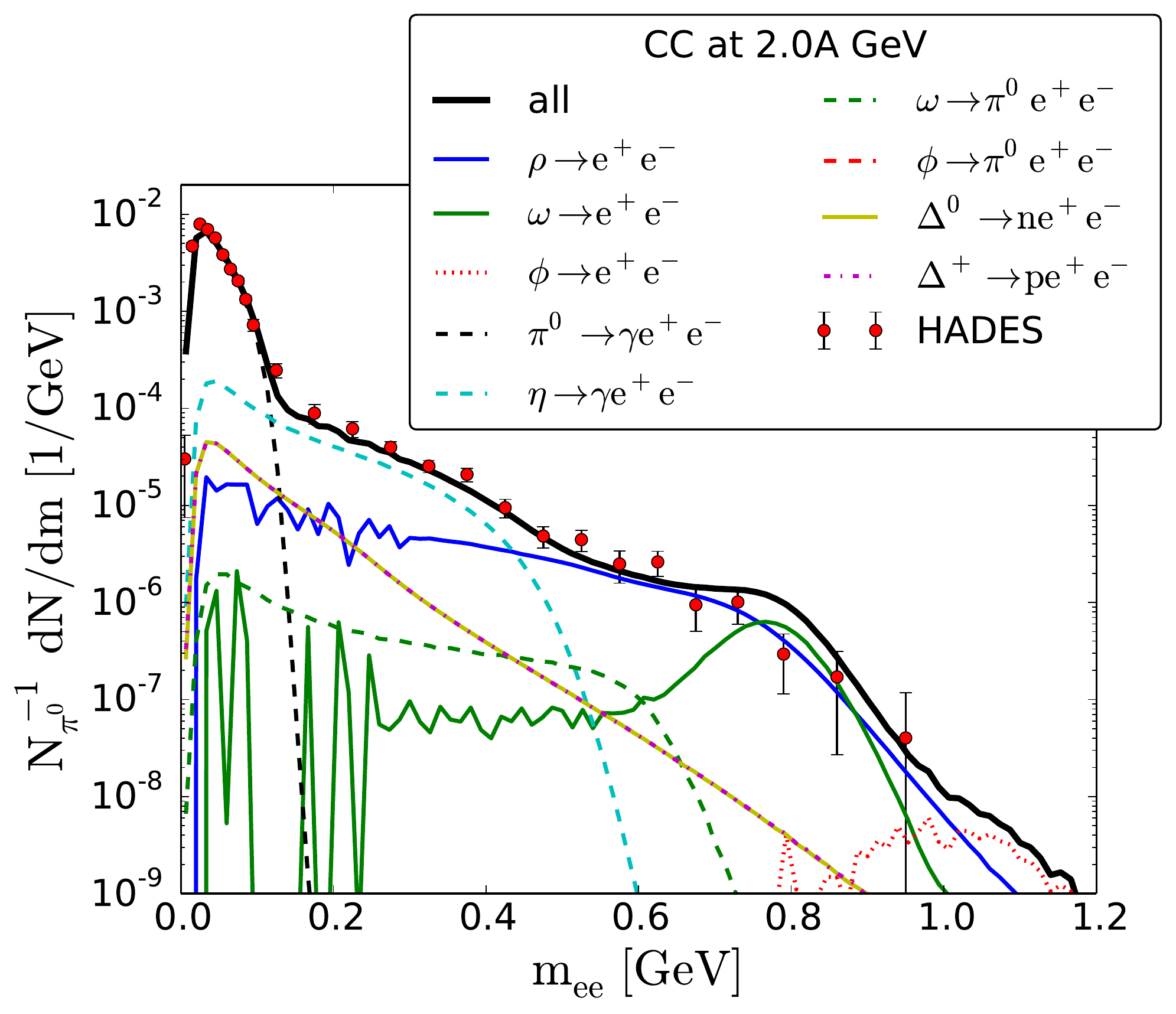}
    \caption{\label{fig:CC2mass} Invariant mass spectrum of dielectrons produced by CC reactions at $E_{\rm{Kin}}=2.0A\,\textrm{GeV}$. Data from \cite{Agakishiev:2009zv}.}
\end{figure}

The results for the dilepton invariant mass spectrum in CC collisions at the higher energy of $E_{\rm{Kin}}=2.0A\,\textrm{GeV}$ are shown in Fig.~\ref{fig:CC2mass}. The $\pi$ and $\eta$ contributions dominate the spectrum up to $0.4$~GeV, while above this mass the yield mainly consists of $\rho$ and $\omega$ contributions. At the highest masses, the $\phi$ peak is broadened due to the low resolution of the detector. The data \cite{Agakichiev:2006tg} are nicely described by the total yield. In the region around the $\omega$-pole mass ($M_{0,\omega} = 0.783$ GeV), the $\rho$ and $\omega$ contributions are slightly overestimated, which might already indicate an onset of in-medium modifications. Following the same argumentation as for $E_{\rm{Kin}}=1.0A\,\textrm{GeV}$, that CC equals a NN superposition as reported in~\cite{Agakishiev:2009yf}, the agreement with the experimental spectrum  suggests that the dilepton emission for elementary np collisions for energies higher than $E_{\rm{Kin}}=1.0\,\textrm{GeV}$ is in better agreement with experimental data.

Compared to the most recent results from the similar UrQMD transport approach for the same system (\cite{Endres:2013nfa}, Figure 1), the results presented here compare overall similarly to data, but are in better agreement in the low and the vector meson pole-mass region. That the agreement is similar or better holds also for the results from UrQMD for other systems reported in \cite{Endres:2013nfa,Schmidt:2008hm}. On the one hand, differences originate in the detailed investigation of the $\rho$-like contribution by studying the decay of baryonic resonances (Fig.~\ref{fig:pp3.3origin}). The input of the relevant branching ratios in SMASH was carefully constrained by the dilepton data and more recent PDG data~\cite{Agashe:2014kda}. On the other hand, the different thresholds of the vector meson contributions of the $\rho$ and the $\omega$ lead to notable differences. In particular for the here discussed CC system, the $\rho$ is, as the second largest yield, a significant contribution in the low-mass region. The $\omega$ low-mass tail is not important for the overall yield. The relative difference to the $\omega$ Dalitz decay is, however, smaller than for pp. In principle, the $\phi$ meson again has low-mass contributions, but its yield is again too small to be visible on the chosen scale. 

The fact that the description of the dilepton production with SMASH matches the data for $E_{\rm{Kin}}=2.0A\,\textrm{GeV}$ validates the resonance treatment and the approach for this energy. It also shows that no in-medium modifications seem to be necessary to describe the dilepton production for such small systems or at least that invariant mass dilepton data are not sensitive to such modifications. It is important to mention, however, that transport approaches include collisional broadening, even if the description of the resonances is still based on vacuum properties.

\subsubsection{ArKCl} \label{sec:arkcl}

On the basis of the dilepton production in elementary and small nucleus-nucleus systems shown above, larger systems are explored. A good example for an intermediate-sized collision system is the ArKCl system at $E_{\rm{Kin}}=1.76A\,\textrm{GeV}$ measured by HADES \cite{Agakishiev:2011vf}. Within SMASH it is modeled with a $^{40}\rm{Ar}$ projectile hitting a $^{37}\rm{Ar}$ nucleus target to emulate an average of the $^{35}\rm{Cl}$ and $^{39}\rm{K}$ composition.

\begin{figure}
    \includegraphics[width=0.95\columnwidth]{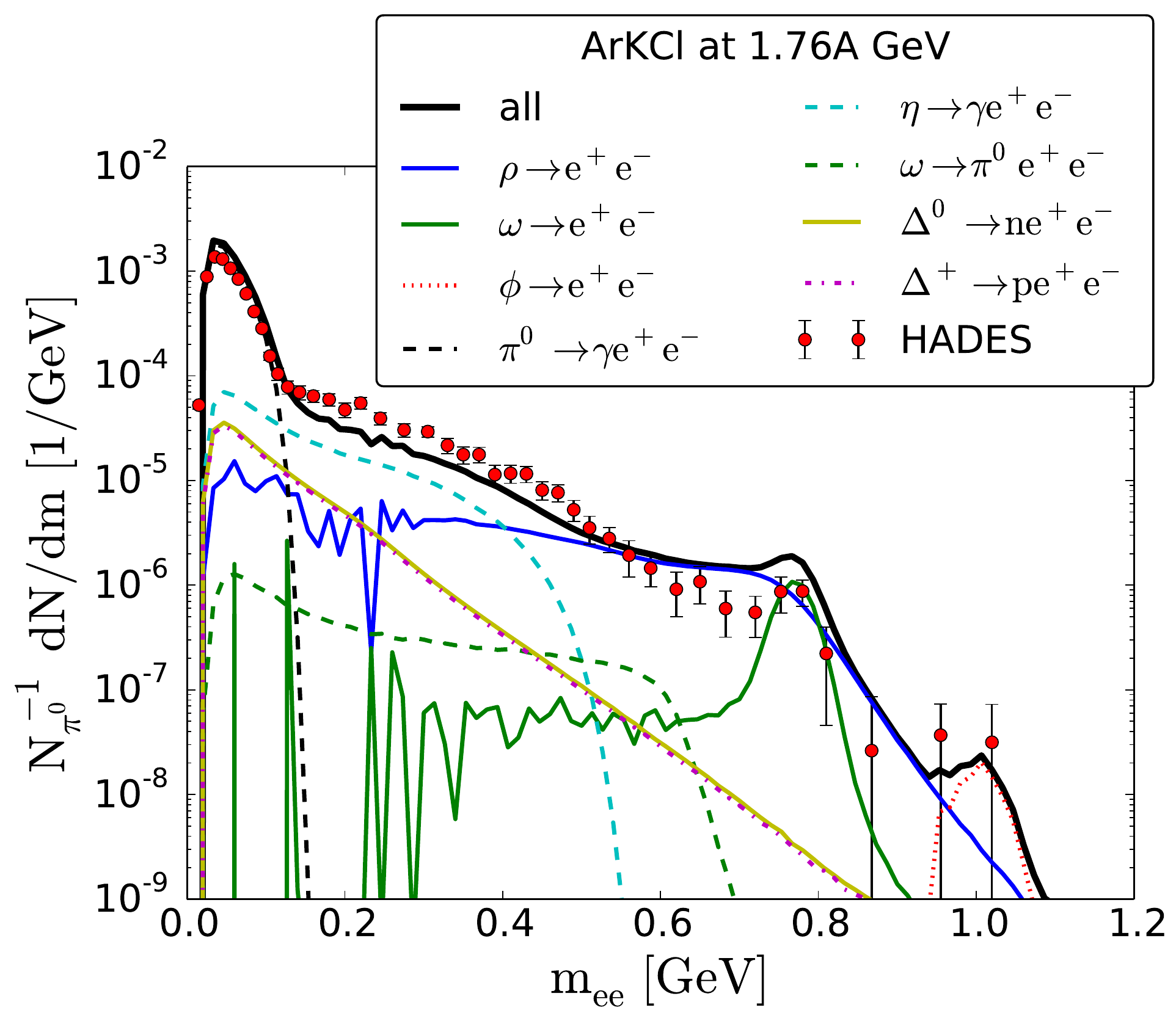}
    \caption{\label{fig:ArKClmass} Invariant mass spectrum of dielectrons produced by ArKCl collisions at $E_{\rm{Kin}}=1.76A\,\textrm{GeV}$. Experimental data from \cite{Agakishiev:2011vf}.}
\end{figure}

Like most of the discussed dielectron invariant mass spectra, the ArKCl yield shown in Fig.~\ref{fig:ArKClmass} is dominated by $\pi$ and $\eta$ in the low and the vector mesons ($\rho$, $\omega$, $\phi$) in the higher invariant mass region above $0.5$~GeV. Since there are more neutrons than protons in the colliding nuclei, the $\Delta^0$ is slightly above the $\Delta^+$ yield. Although the spectrum is in reasonable agreement with the experimental data \cite{Agakishiev:2011vf} for low and highest invariant masses, two distinct issues are revealed by the comparison to experimental data. First, the $\rho$ contribution is too large in the region between $0.6$~GeV and $0.8$~GeV. This might be connected to the overproduction in the exclusive $\rho$ cross section discussed in Sec.~\ref{sec:xs}. The $\rho$ in pp reactions at $E_{\rm{Kin}}=1.76A\,\textrm{GeV}$ ($\sqrt{s_{NN}}=2.61\,\textrm{GeV}$) is almost solely produced by the overestimated exclusive process $pp\rightarrow pp\rho$ (Fig.~\ref{fig:xs_rho}). The second issue is an underestimation in the mass region between $0.15$~GeV and $0.5$~GeV.

In combination both issues indicate the limit of the assumption of resonances with vacuum properties. The low mass region is known to be enhanced by in-medium modifications, i.e. a broadening of the vector meson spectral functions, in particular of the $\rho$ meson \cite{Rapp:2009yu}. This broadening has also an influence on the $\rho$ pole mass region, since the yield in this region will be decreased by a broadening. That in-medium modifications are relevant is also supported by the agreement with experimental data for CC reactions at the similar energy of $E_{\rm{Kin}}=2.0A\,\textrm{GeV}$. The agreement is expected to translate to a larger system, if no additional medium effects become relevant.

This result therefore suggests that dilepton emission in systems as large as ArKCl is sensitive to in-medium modifications of resonances that go beyond the intrinsic collisional broadening mentioned above. To verify this hypothesis, a comparison with a coarse-graining approach \cite{Endres:2015fna} that employs in-medium modifications of spectral functions is presented in Sec.~\ref{sec:cg}.

\subsubsection{AuAu} \label{sec:auau}

The largest collision system discussed in this work is gold-gold (AuAu) scattering at $E_{\rm{Kin}}=1.23A\,\textrm{GeV}$, matching upcoming HADES results \cite{Franco:2017ano}.

\begin{figure}
    \includegraphics[width=0.95\columnwidth]{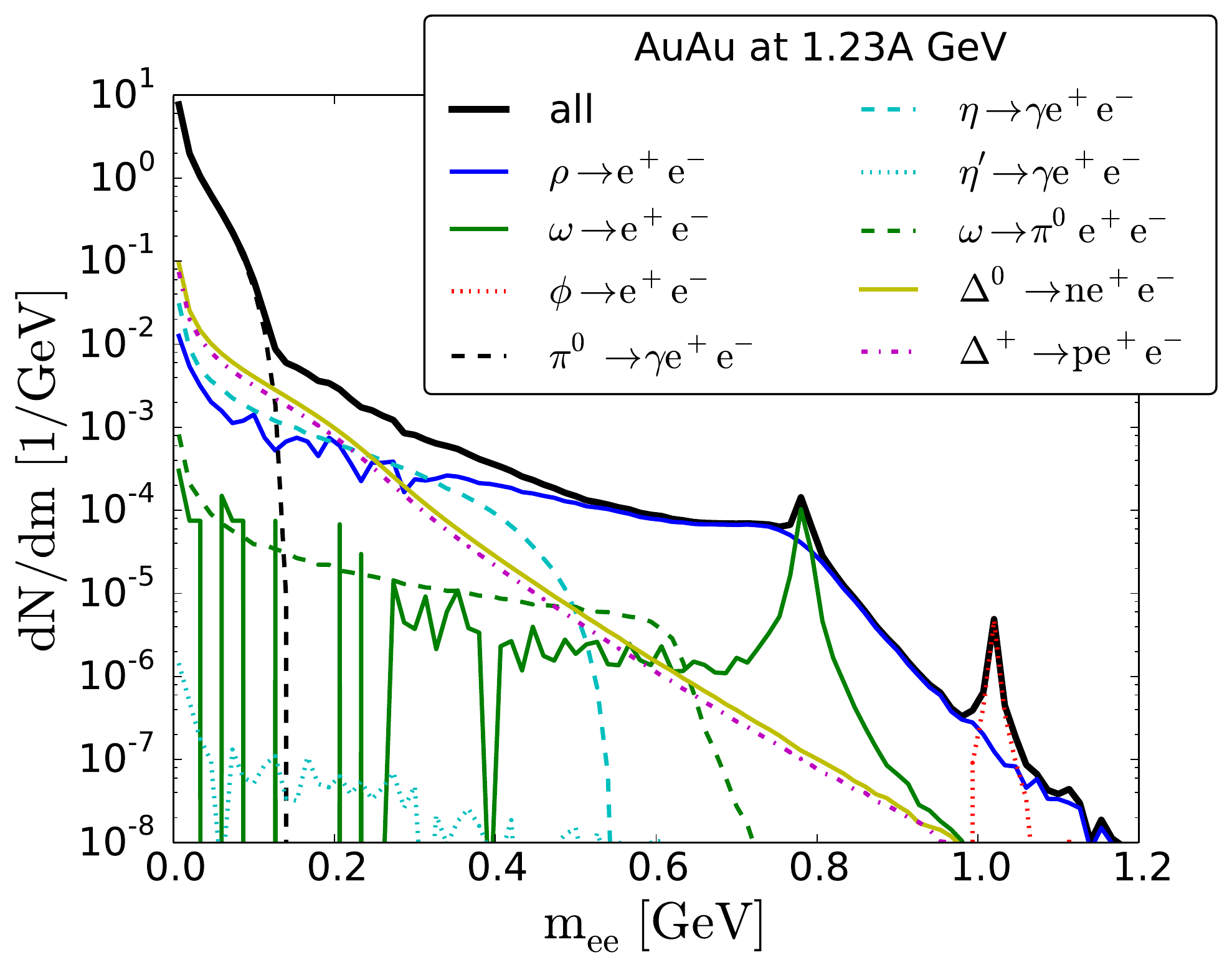}
    \caption{\label{fig:AuAumass} Invariant mass spectrum of dielectrons produced by AuAu collisions at $E_{\rm{Kin}}=1.23A\,\textrm{GeV}$.}
\end{figure}

The invariant mass dielectron spectrum (Fig.~\ref{fig:AuAumass}), which is not acceptance filtered, reveals some differences to the previous cases even without a comparison to experimental data. The $\Delta$ yield is larger in relation to other contributions than for smaller systems. The $\phi$ peak is most prominent in this spectrum and the $\rho$ contribution shows a slight bump at the pole mass, since the reaction $\pi\pi\rightarrow\rho$ dominates over the different Dalitz-like $N^*$ and $\Delta^*$ contributions. Both effects can be explained by the large amount of secondary reactions. Fig.~\ref{fig:AuAumass} also shows an $\eta'$ contribution, which is only visible since a large vertical scale is chosen for this plot. This illustrates that its contribution is negligible, especially for smaller systems. Furthermore, the limited statistics also suggests that $\eta'$ are produced rarely even in a large system. 

The $\rho$ is dominant up to the $2m_{\pi}$ threshold and remains one of the leading contributions for lower masses. Even though the statistics is limited below the hadronic threshold, it can be seen that the direct $\omega$ contribution is on the same order of magnitude as the $\omega$ Dalitz contribution. This shows that towards larger systems (compare Fig.~\ref{fig:pp3.5mass} and Fig.~\ref{fig:CC2mass}) the difference between the direct and Dalitz $\omega$ decay becomes smaller. In other words, the sub-threshold contributions become more prominent the larger the system. Only the $\phi$ contribution remains small, since it is suppressed for the here discussed low energies. This might change for higher energies with a larger overall $\phi$ production.

Extrapolating from the already performed experimental data comparisons, an even larger underestimation in the intermediate mass region, due to the greater importance of in-medium modifications in a larger medium, and an overestimation in the $\rho$ pole mass region is expected for a future experimental data comparison. New data will be valuable to further constrain e.g. the $\phi$ production, as well as clarifying the role of medium effects.

\subsection{Coarse-Graining Approach \label{sec:cg}}

In order to investigate the effect of in-medium modifications, the hadronic evolution of SMASH is coarse grained (CG) in this section following the original idea from \cite{Huovinen:2002im}. This means that macroscopic quantities are extracted locally from the microscopic transport model, enabling the determination of thermal dilepton emission from those regions. The dilepton radiation is therefore a mix of dilepton production from thermal dilepton emission rates and the usual hadronic transport contributions. The framework used for this work is the same as in \cite{Endres:2015fna} and has proven to be reliable in describing experimental data from SIS up to LHC energies \cite{Endres:2015fna, Endres:2014zua, Endres:2016tkg, Galatyuk:2015pkq}. In the following, a brief summary of the approach is given, for a comprehensive review the reader is referred to \cite{Endres:2015fna}. The approach locally averages over the reaction evolution by splitting an ensemble of collision events into small space-time cells. For those cells the baryon density $\rho_B$ and the energy density $\epsilon$ are calculated in the rest-frame. Knowing both and assuming (for the beam energies considered in this work) a hadron resonance gas equation of state, the local temperature $T$ and baryon chemical potential $\mu_B$ are determined. Based on the thermodynamic information, the yield of dileptons from a certain cell is given by the corresponding thermal emission rates that include the in-medium modification on the vector meson spectral function. The in-medium description used in the coarse-grained approach employed here is based on the hadronic many-body theory \cite{Rapp:1999us, Rapp:2000pe}, where the spectral function depends on temperature and density. Since in-medium modifications at these energies are only expected to affect the $\rho$ and $\omega$ significantly, dilepton yields from thermal rates are only calculated for these two. If the temperature drops inside the cells, the assumption of thermal rates is no longer reasonable; therefore also non-thermal (\emph{freeze-out}) transport contributions are included, which are known to be of significance only around the pole masses~\cite{Endres:2015fna}. The $\omega$ Dalitz decay is also part of the  $\omega$ freeze-out contribution. Thermal rates together with the freeze-out contribution form the coarse-graining contributions for the $\rho$ and $\omega$ (CG-$\rho$ and CG-$\omega$). The last contribution from the coarse-graining approach are multi-$\pi$ states originating from broad resonances. The dilepton cocktail is completed with the relevant transport contributions of $\pi$, $\eta$ and $\phi$ from SMASH.

\begin{figure}
    \includegraphics[width=0.96\columnwidth]{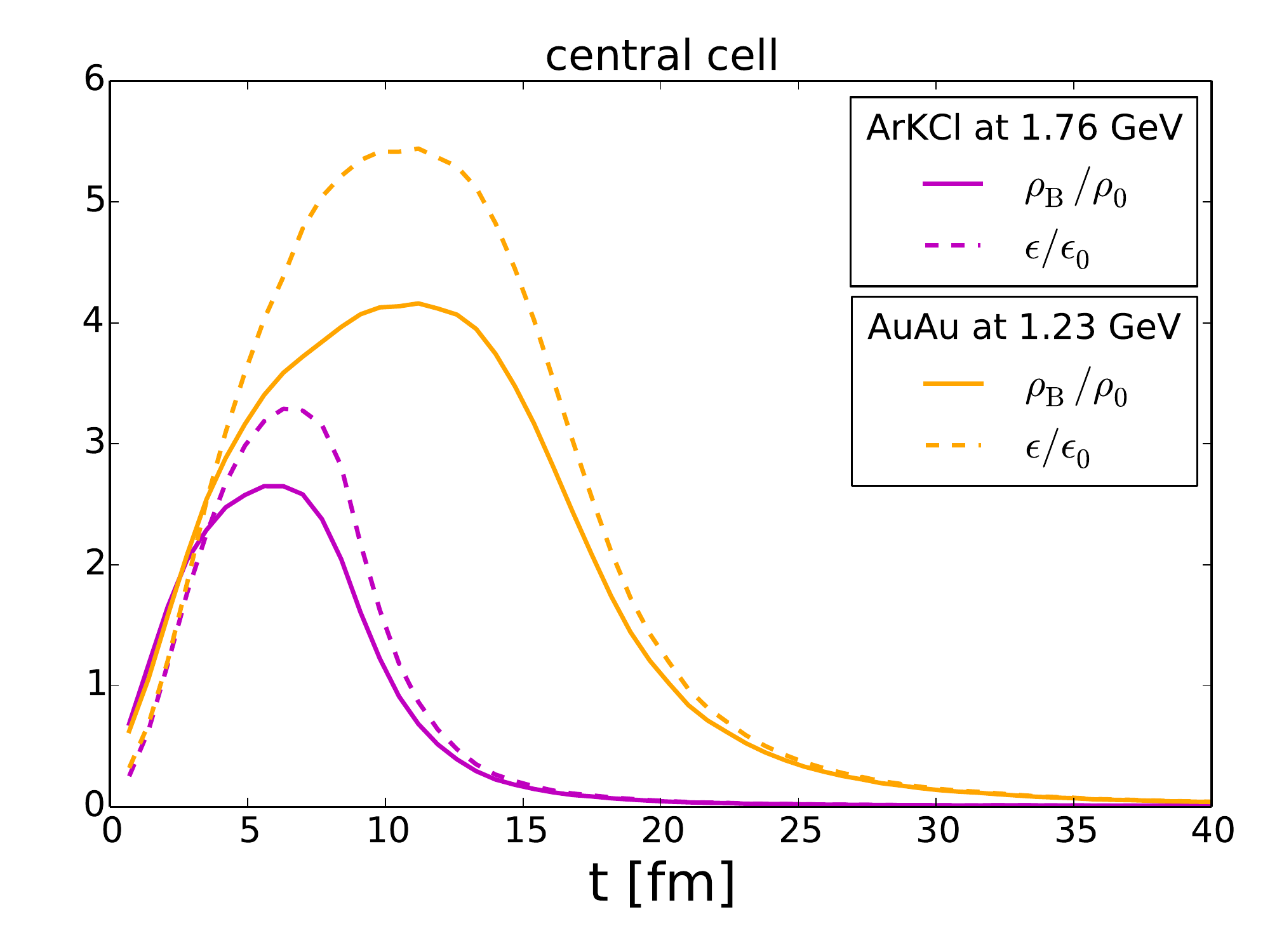}
    \caption{\label{fig:dens_evo} Evolution of the energy and baryon density in units of the ground state density in the most central cell over time for ArKCl collisions at $E_{\rm{Kin}}=1.76A\,\textrm{GeV}$ and AuAu collisions at $E_{\rm{Kin}}=1.23A\,\textrm{GeV}$.}

    \includegraphics[width=0.96\columnwidth]{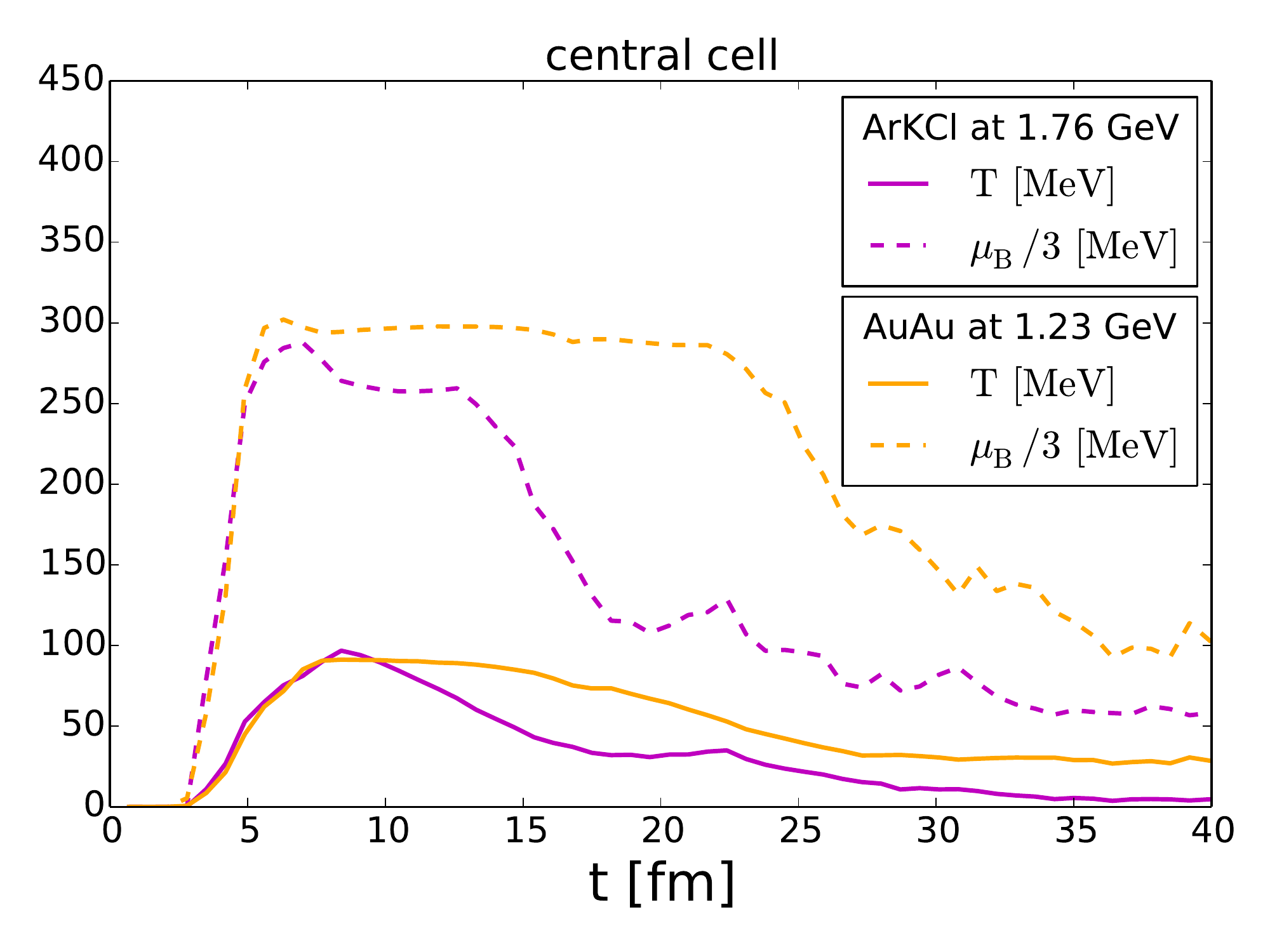}
    \caption{\label{fig:temp_evo} Evolution of the temperature $T$ and baryochemical potential $\mu_b$ in the most central cell over time for ArKCl collisions at $E_{\rm{Kin}}=1.76A\,\textrm{GeV}$ and AuAu collisions at $E_{\rm{Kin}}=1.23A\,\textrm{GeV}$.}
\end{figure}

Before looking at the results for dilepton emission in the context of the experimental data, let us show the thermodynamic properties of the system. An example for the evolution of the baryon $\rho_B$ and energy $\epsilon$ density in units of the ground-state density in ArKCl and AuAu collisions at SIS energies is given in Fig.~\ref{fig:dens_evo} for the central cell at the origin of the coarse-graining grid. As expected, initially the density rises similarly for both studied systems, while for the AuAu collisions both densities rise higher and reach a larger maximum ($5\times$ ground state density) later. For both systems the density falls with time, but faster for the smaller ArKCl system.

Fig.~\ref{fig:temp_evo} shows the temperature and baryon chemical potential for the same systems and central cell as above. The baryon chemical potential quickly rises to a value of around $900$~MeV (Fig.~\ref{fig:temp_evo} shows $\mu_B/3$) at $5$ fm for both systems. The temperature reaches its maximum of $100$~MeV at around $8$ fm and is similar for ArKCl and AuAu. In this case, the plateau in the chemical potential for AuAu even extends from $5$ to around $22$ fm. Note that the values shown here are maximum values at the point of highest density in the system to indicate the reach in the phase diagram.

In summary, the presented results confirm that the cell evolution is reasonable, since the expectation as well as results reported in \cite{Endres:2015fna, Seck:2017opx} that are based on hadronic space-time evolution of UrQMD are matched. This not only further validates the SMASH approach, but forms the basis for the more advanced analysis of the dilepton emission of the coarse-grained evolution.

Results for two large systems are presented in the following - ArKCl and AuAu. The two different approaches used in this work are compared in the following: First, the dilepton yield from the transport model SMASH as discussed in the previous sections (referred to as \emph{non-CG}); although the medium effect of collisional broadening is included, no in-medium modifications are incorporated. Second, the outcome from the coarse-graining approach, which employs thermal rates including an in-medium description for the $\rho$ and $\omega$ meson; those medium-modified dilepton contributions are combined with unmodified cocktail contributions ($\pi$, $\eta$, $\phi$) from the SMASH simulations.


\subsubsection{Results for the dilepton emission}

\begin{figure}
    \includegraphics[width=0.96\columnwidth]{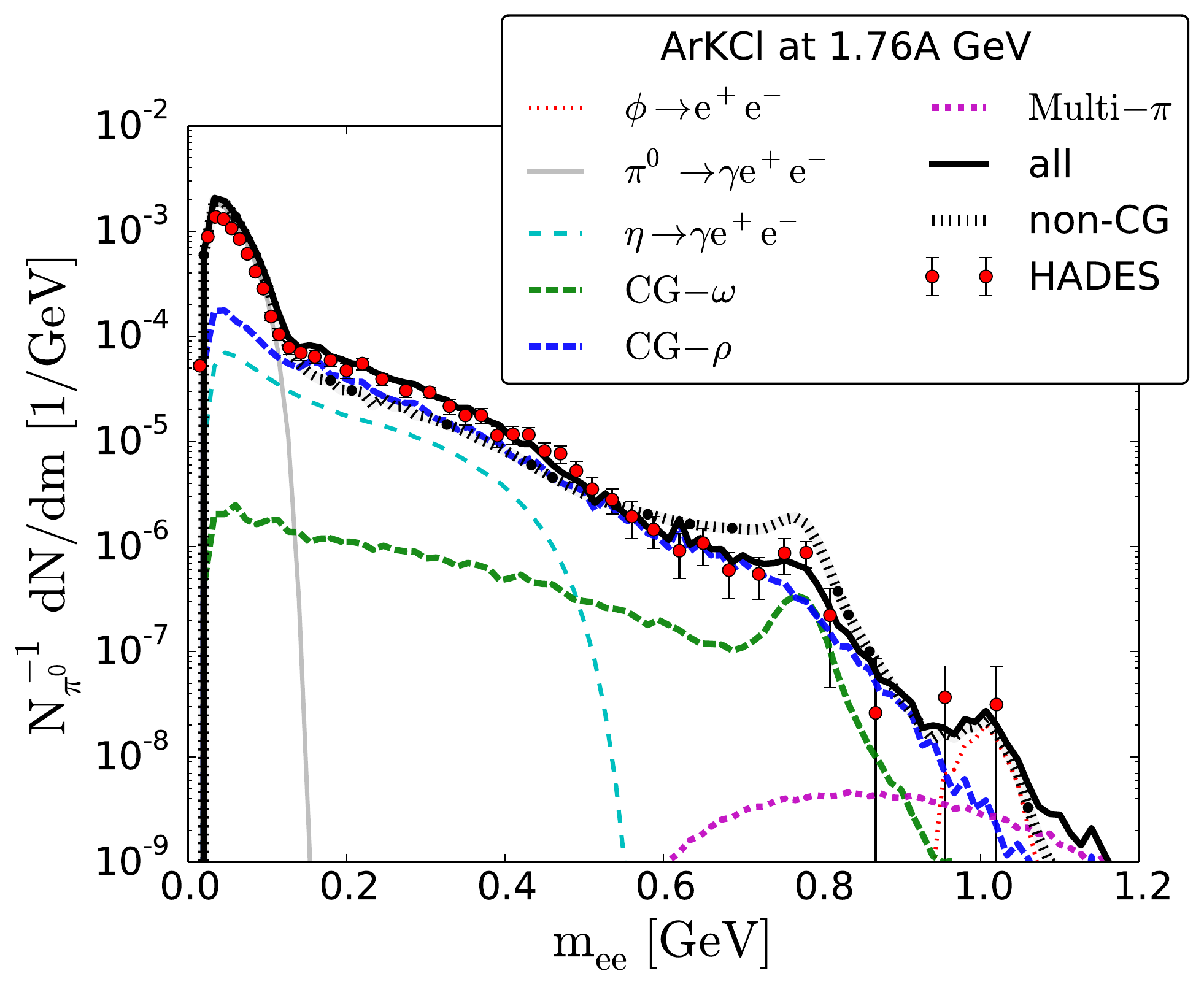}
    \caption{\label{fig:CGArKCl} Invariant mass spectrum of dielectrons produced by ArKCl collisions at $E_{\rm{Kin}}=1.76A\,\textrm{GeV}$ within the Coarse-Graining approach. Dashed lines from coarse graining and solid lines from SMASH dilepton production (as in Fig.~\ref{fig:ArKClmass}). Experimental data from \cite{Agakishiev:2011vf}.}
\end{figure}

First, the focus is on the dilepton emission from ArKCl collisions at $E_{\rm{Kin}}=1.76A\,\textrm{GeV}$. The total SMASH vacuum transport result (\emph{non-CG}) is underestimating the invariant mass spectrum for this system in the low mass region and overestimating it in the $\rho$ pole mass region (as discussed in Sec.~\ref{sec:arkcl}). Fig.~\ref{fig:CGArKCl} shows the results from the coarse-graining approach, which again are cut in momentum and angular distribution as well as filtered for the acceptance \cite{Janus:HAFT} in order to compare to experimental data from HADES \cite{Agakishiev:2011vf}. All solid lines refer to SMASH dilepton production (the same as in Fig.~\ref{fig:ArKClmass}). Out of these only the $\pi$ and $\phi$ yields are important for the overall spectrum at low and high invariant masses respectively. The dashed contributions for $\rho$ and $\omega$ display results from the coarse-graining approach and include the thermal dilepton rates containing in-medium modifications and the freeze-out contributions for cold cells. Also, the multi-$\pi$ contribution is added, but has only little effect on the overall spectrum due to the low beam energy. On the contrary, the effects on the vector meson dilepton yield are large. For both, $\rho$ and $\omega$, the yield is shifted from the pole mass to the low-mass region.

Quantitatively, the agreement of the overall yield (\emph{all}) with experimental data \cite{Agakishiev:2011vf} in Fig.~\ref{fig:CGArKCl} is greatly improved with the in-medium modifications of the vector meson spectral functions employed in the coarse-graining approach compared to the SMASH dilepton production based on vacuum resonance properties. Only the normalization on the $\pi$ multiplicity leads to an overestimation of the $\pi$ peak for low invariant masses and consequently also to a slight overproduction around $0.15$~GeV. It can be concluded that a sensitivity to in-medium modifications in the ArKCl spectrum is confirmed within our approach.


\begin{figure}
    \includegraphics[width=0.96\columnwidth]{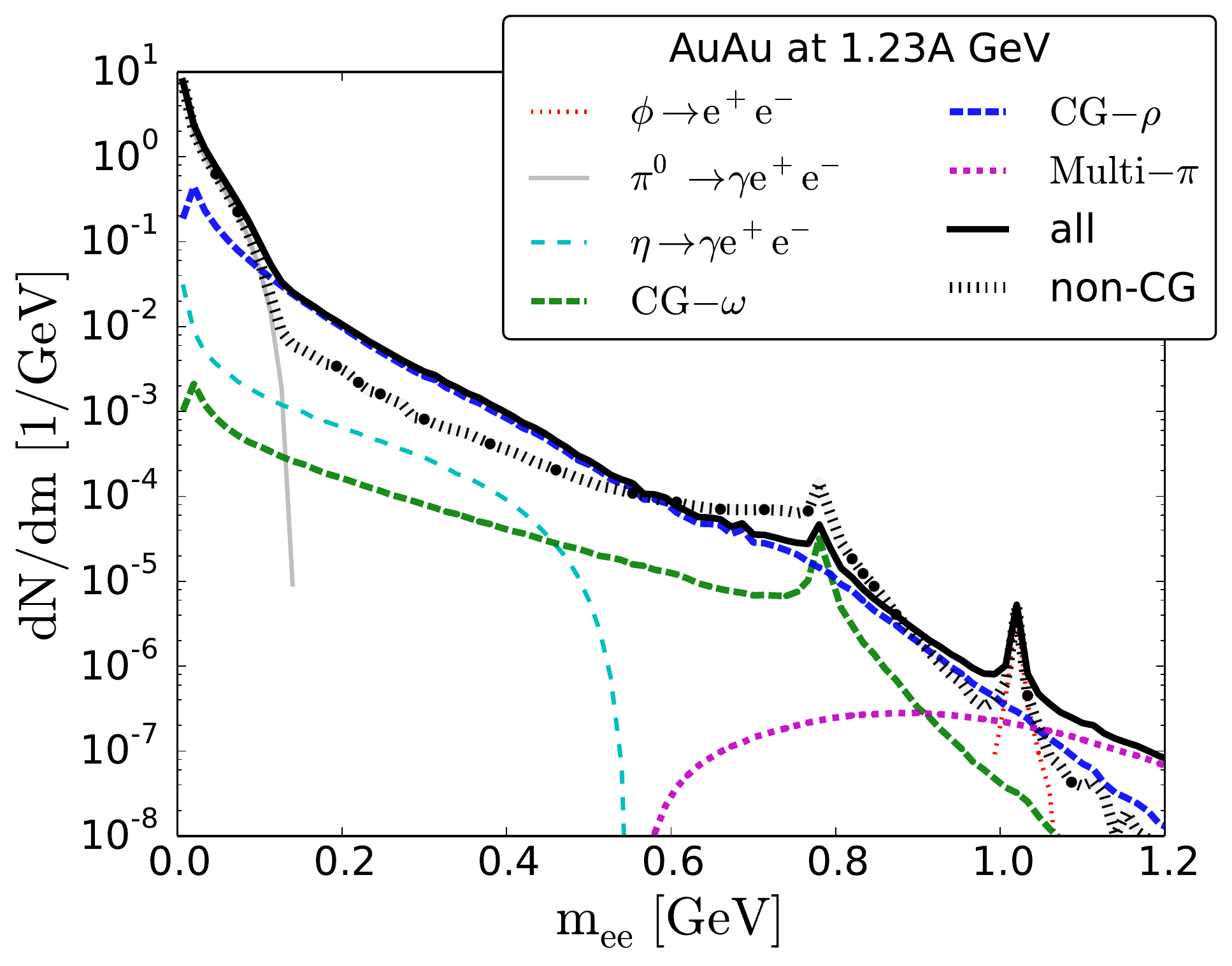}
    \caption{\label{fig:CGAuAu} Invariant mass spectrum of dielectrons produced by AuAu collisions at $E_{\rm{Kin}}=1.23A\,\textrm{GeV}$ within the Coarse-Graining approach. Dashed lines from coarse graining and solid lines from SMASH dilepton production (as in Fig.~\ref{fig:AuAumass}).}
\end{figure}

For the larger AuAu system the effect of in-medium modifications is expected to be larger than in ArKCl collisions. Predictions for AuAu collisions at $E_{\rm{Kin}}=1.23A\,\textrm{GeV}$ within the coarse-graining approach, which are not acceptance filtered, are shown in Fig.~\ref{fig:CGAuAu}. Even though no experimental data constraints are available yet, the comparison of the SMASH dilepton result for the total yield from Fig.~\ref{fig:AuAumass} (\emph{non-CG}) with the spectrum from the coarse-graining approach (\emph{all}) already hints at a larger modification of the yield. Differences are observed in the low-mass region and around vector meson pole masses analogously to ArKCl, but the effect in the intermediate-mass region seems more pronounced. Also in the intermediate invariant mass region the multi-$\pi$ contribution leads to deviations. Again the only relevant yields from the SMASH contributions are the $\pi$ and $\phi$ channel.

The findings for both ArKCl and AuAu align nicely with previous results obtained with the UrQMD coarse-graining transport approaches \cite{Endres:2015fna, Seck:2017opx}. 

\subsubsection{Comparison of the vector meson yields} \label{sec:cg_comp}

The study presented here allows for a unique direct comparison between a coarse-graining approach and the dilepton production from a hadronic transport approach. This becomes possible since both rely on the same hadronic evolution. SMASH also includes low-mass vector meson contributions, which enable a comparison in this mass region. The comparison for the total yield was already shown in the previous section. This section focuses specifically on the vector meson ($\rho$ and $\omega$) dilepton contributions, which are of most interest, since for them an in-medium spectral function is employed. Therefore, this allows to contrast the different medium effects at play: On the one side, the here employed transport approach with collisional broadening and with vacuum resonance properties is shown, which includes the coupling of baryonic resonances to the vector mesons via Dalitz decays (as discussed in Sec.~\ref{sec:pp}). On the other side is a full in-medium description of the vector meson spectral functions.

\begin{figure}
    \includegraphics[width=0.96\columnwidth]{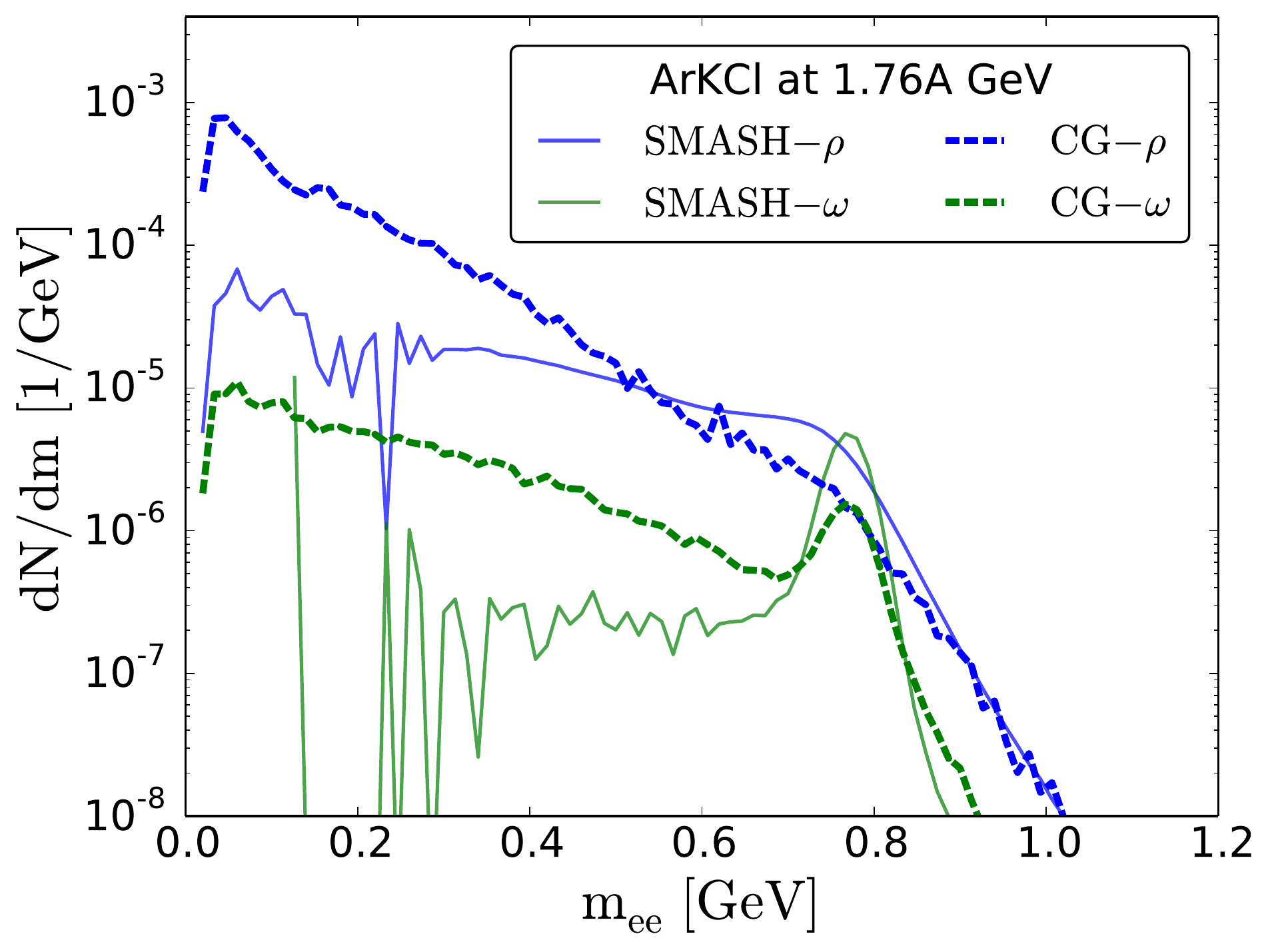}
    \caption{\label{fig:CGArKCl_res} Comparison of invariant mass spectra of dielectrons produced by $\rho$ and $\omega$ in ArKCl collisions at $E_{\rm{Kin}}=1.76A\,\textrm{GeV}$ within the Coarse-Graining approach versus the default SMASH dilepton production.}
\end{figure}

Fig.~\ref{fig:CGArKCl_res} shows the direct comparison between the vector meson yields for the ArKCl system discussed above. As expected from a broadening scenario of the vector meson spectral function, an enhancement of the low-mass tail together with a decrease in the pole mass region, especially for the $\rho$ yield, is observed. The $\rho$ yield from the coarse graining shows an almost exponential decrease with mass and dominates the spectrum over most of the covered invariant mass range. The $\omega$ yield is only around its pole mass equally large. Also, the broadening of the $\omega$ spectral function is nicely reflected in the dilepton emission around the peak.

\begin{figure}
    \includegraphics[width=0.96\columnwidth]{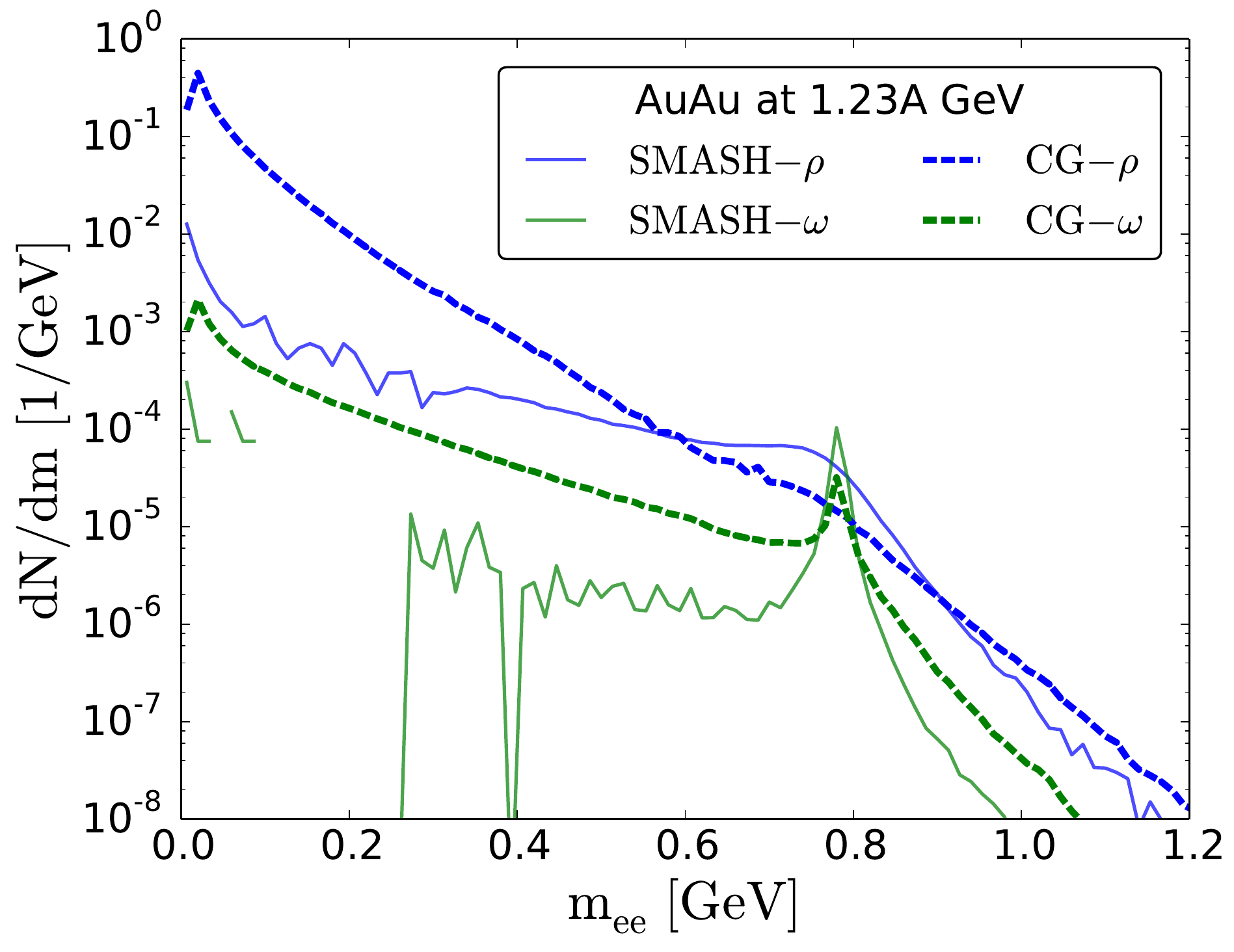}
    \caption{\label{fig:CGAuAu_res} Comparison of invariant mass spectra of dielectrons produced by $\rho$ and $\omega$ in AuAu collisions at $E_{\rm{Kin}}=1.23A\,\textrm{GeV}$ within the Coarse-Graining approach versus the default SMASH dilepton production.}
\end{figure}

Fig.~\ref{fig:CGAuAu_res} again shows the shift away from the pole masses in the $\rho$ and $\omega$ yield, in this case for the AuAu system. Again, the broadening of the $\omega$ peak and the enhancement of the $\rho$ low-mass tail is nicely observed. The shape of the $\rho$ contribution almost completely flattens out and reveals an exponential decrease, whereas for the $\omega$ a dominating peak around the pole mass remains.

Both comparisons in Fig.~\ref{fig:CGArKCl_res} and Fig.~\ref{fig:CGAuAu_res} interestingly reveal that the SMASH contributions obtain a low-mass tail similar to the CG contributions. In SMASH, the Dalitz-like tail for low masses stems from the different baryonic-resonance contributions (compare Fig.~\ref{fig:pp3.3origin}). Although this is not a medium effect, since it is already observed in proton-proton reactions, the underlying mechanism leading to the pronounced low-mass tail, namely the coupling of the vector meson to baryonic resonances, is the same that is found to be important for the in-medium modifications of the spectral functions used in the coarse graining framework \cite{Rapp:1999us, Rapp:2000pe}. It is clear from the differences that collisional broadening plus baryonic Dalitz decays cannot account for the whole effect of the medium. Nevertheless, the presented results quantify the different effects on the dilepton spectrum for the first time.

\section{Summary and Outlook} \label{sec:sum}

In this work, the complete set of available dielectron production measurements at SIS energies based on a new hadronic transport approach (SMASH) is discussed. SMASH relies on resonance interactions with vacuum properties and the corresponding dilepton production is validated by a good agreement with experimental data in proton-nucleus and nucleus-nucleus collisions up to a system size of CC reactions. The agreement originates in the solid description of elementary pp collisions. Only for a low kinetic energy around $1$ GeV is an underestimation for quasi-free np collisions and subsequently for CC is observed. Overall, the description of low energy collisions is comparable with similar transport approaches. 

In SMASH the dilepton decays are taken into account for the spectral function calculation for all vector mesons ($\rho$, $\omega$, $\phi$), leading to contributions from their direct dilepton decays down to $2m_e$ below the hadronic threshold. Such contributions are found to be significant for the low-mass region and to become more prominent the larger the collision system is.

The hadronic transport approach is complemented by a coarse-graining approach based on the same hadronic evolution to study the sensitivity of the invariant mass spectrum to in-medium modifications of the vector meson spectral function. Both revealed similar features in their dilepton contributions. Nevertheless, the transport description including the coupling to baryons and collisional broadening cannot account for the necessary significant modifications visible in larger collision systems caused by an in-medium description of the vector meson spectral function. The significance of an in-medium description is noticed already for the here investigated low energy collisions beginning with systems as large as ArKCl.

The presented results include predictions for upcoming HADES results for $\rm{\pi}$p and AuAu reactions. Comparisons to experimental data of the former specifically probe the coupling of the $\rho$ meson to the $N^*(1520)$ baryonic resonance, while the AuAu system is expected to be even more sensitive to in-medium modifications than the already studied ArKCl collisions indicated by the presented prediction of the coarse-graining approach.

Based on this well-understood baseline for the dilepton production from the hadronic sector for low beam energy collisions in the future the dilepton production in intermediate and high beam energy collisions can be addressed. Dilepton emission from intermediate beam energy collisions is one of the promising observables of the CBM experiment at FAIR. Hybrid approaches allow us to explore the high beam energy reactions of RHIC or LHC by combining the dilepton radiation from a hydrodynamic calculation with the emission from a hadronic afterburner, where SMASH can be applied for the non-equilibrium hadronic evolution.

\begin{acknowledgments}
The authors thank H. van Hees, M. Bleicher, D. Oliinychenko and A. Sch\"afer for fruitful discussions and T. Galatyuk for providing the HADES acceptance filters and data tables. Furthermore, we acknowledge R. Rapp for providing the parametrization of the spectral functions. This work was made possible thanks to funding from the Helmholtz Young Investigator Group VH-NG-822 from the Helmholtz Association and GSI, and supported by the Helmholtz International Center for the Facility for Antiproton and Ion Research (HIC for FAIR) within the framework of the Landes-Offensive zur Entwicklung Wissenschaftlich-Oekonomischer Exzellenz (LOEWE) program from the State of Hesse. H.P. and J.S. acknowledge support by the Deutsche Forschungsgemeinschaft (DFG) through the grant CRC-TR 211 ``Strong-interaction matter under extreme conditions''. Computational resources have been provided by the Center for Scientific Computing (CSC) at the Goethe-University of Frankfurt and the Green IT Cube at GSI.
\end{acknowledgments}

\appendix

\section{Transverse momentum and rapidity spectra for pp collisions at $E_{\rm{Kin}}=3.5\,\textrm{GeV}$ \label{sec:apend_pp3.5}}

\begin{figure*}

    \includegraphics[width=0.45\textwidth]{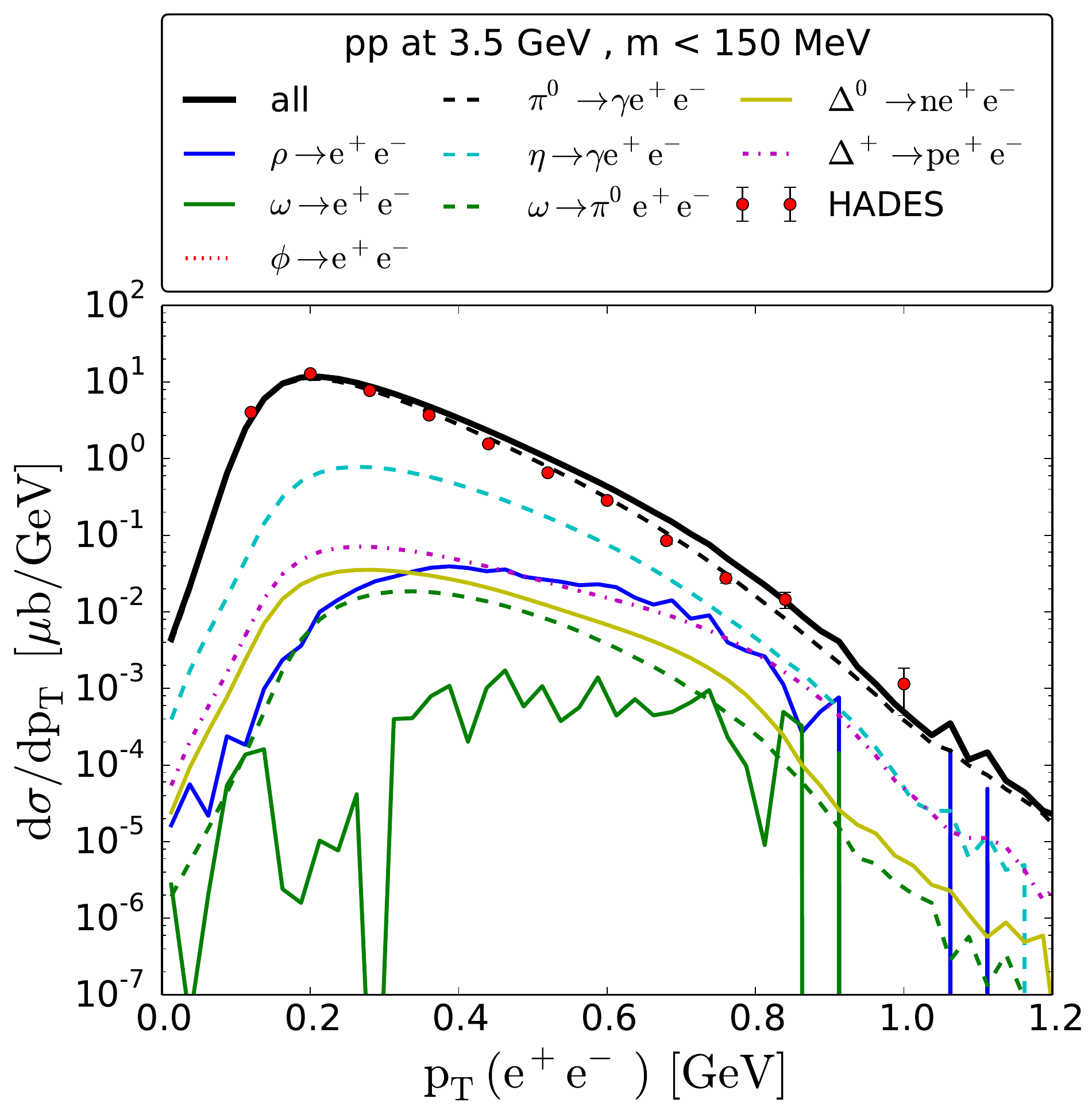}
    \includegraphics[width=0.45\textwidth]{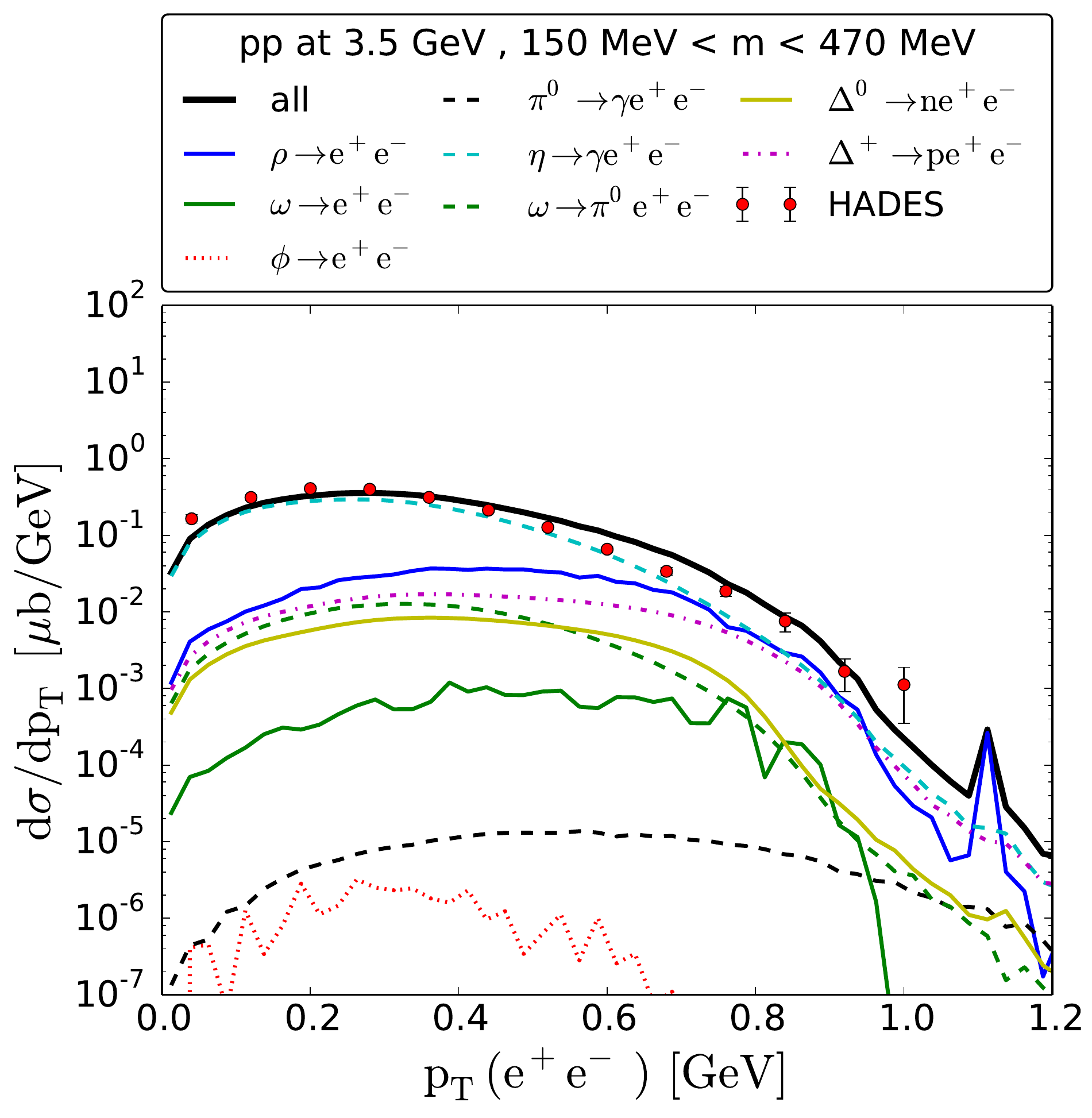}
     \\
    \includegraphics[width=0.45\textwidth]{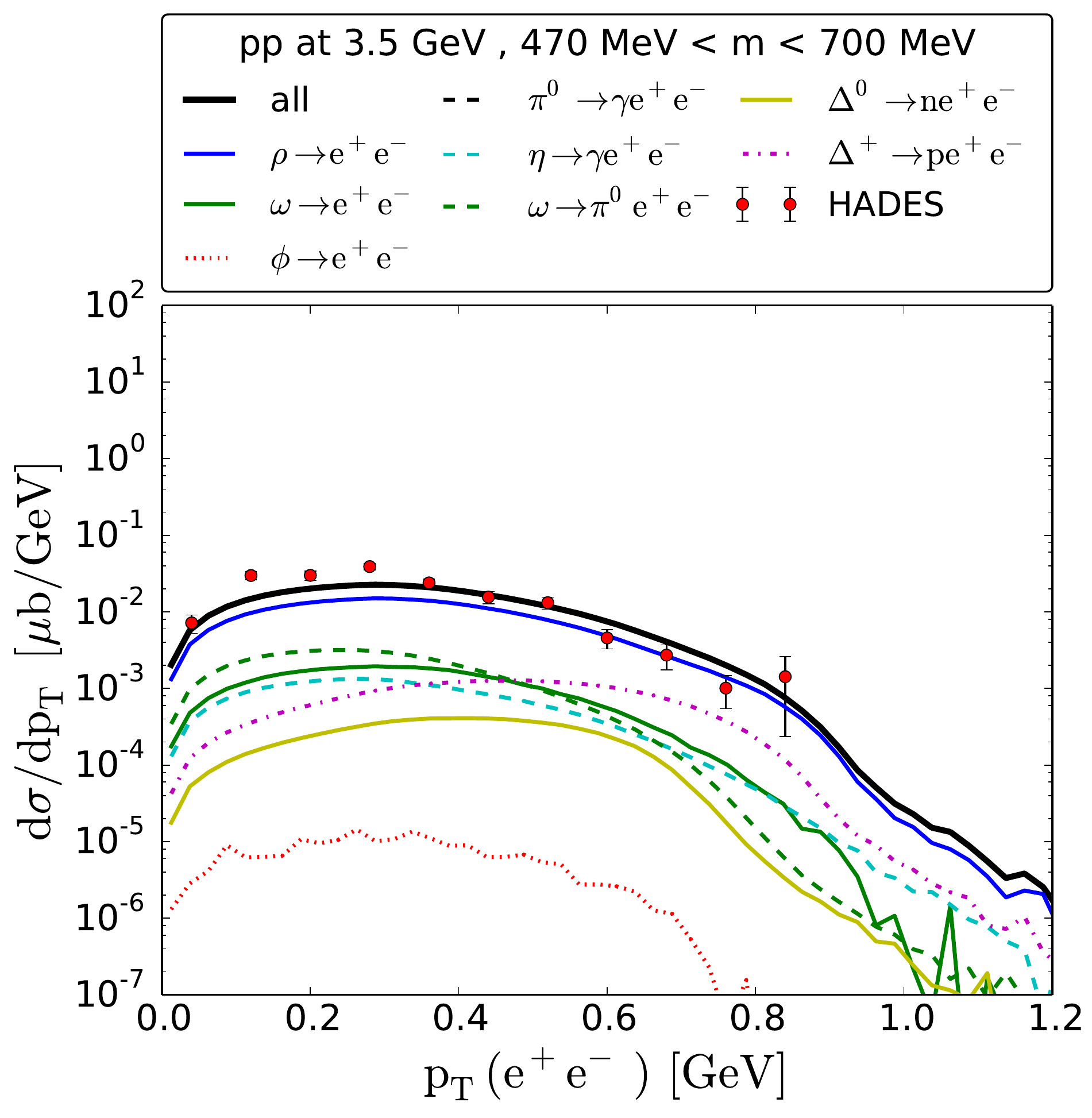}
    \includegraphics[width=0.45\textwidth]{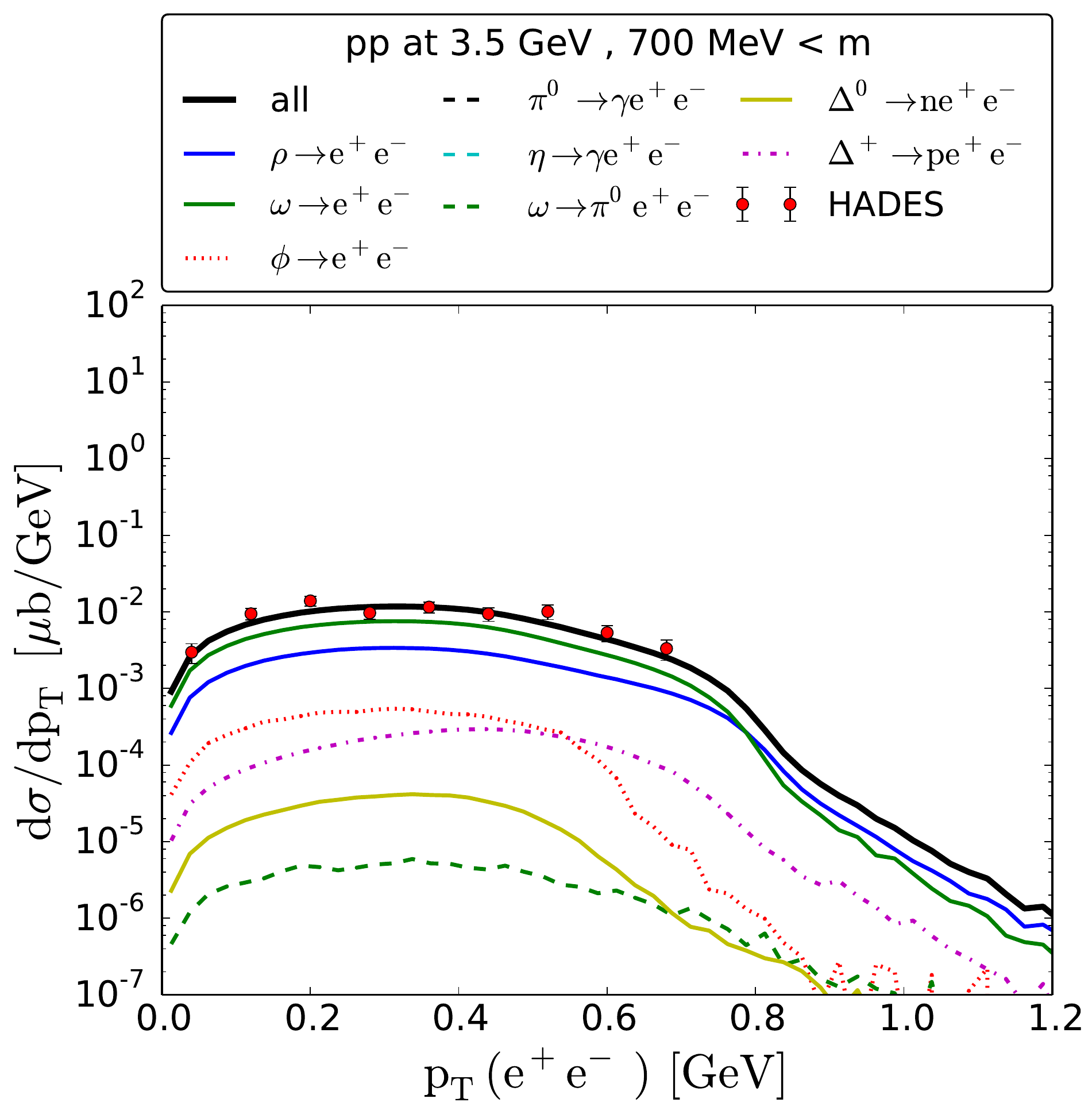}

    \caption{\label{fig:pp3.5pts}Transverse momentum spectra of dielectrons produced by pp collisions at $E_{\rm{Kin}}=3.5\,\textrm{GeV}$ in different invariant mass windows. Experimental data from \cite{HADES:2011ab}.}
\end{figure*}

\begin{figure*}
    \includegraphics[width=0.45\textwidth]{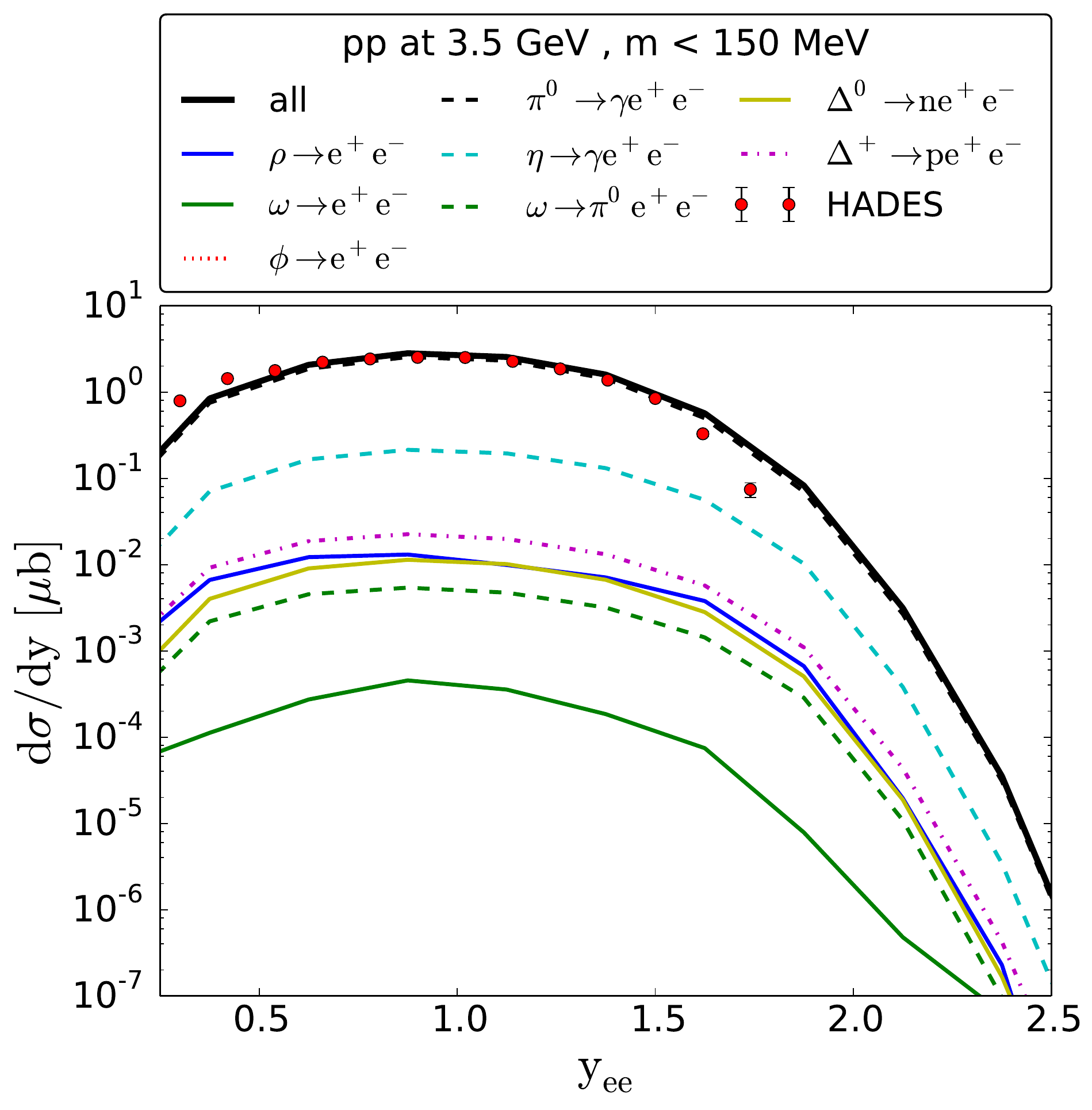}
    \includegraphics[width=0.45\textwidth]{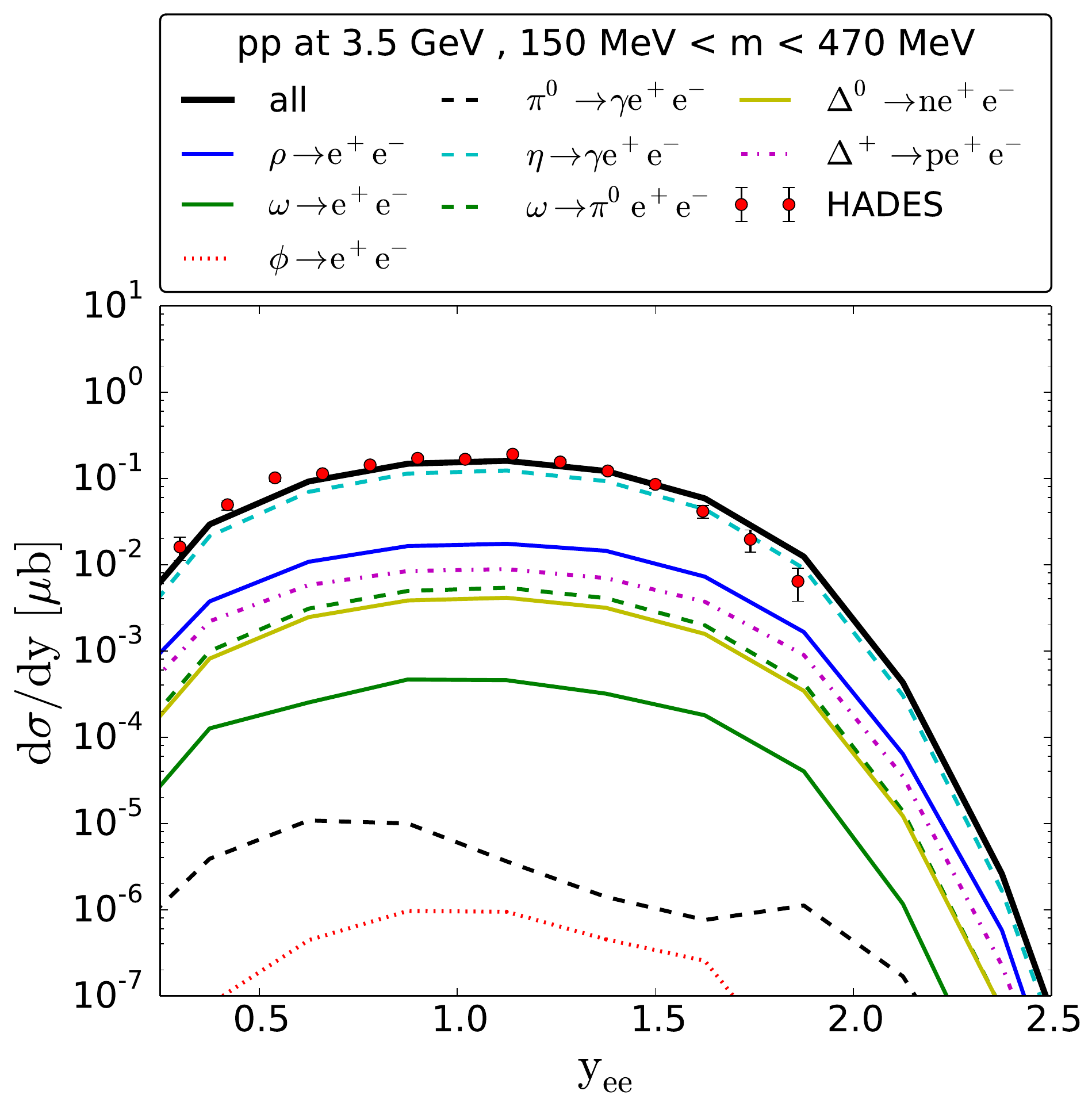} 
    \\
    \includegraphics[width=0.45\textwidth]{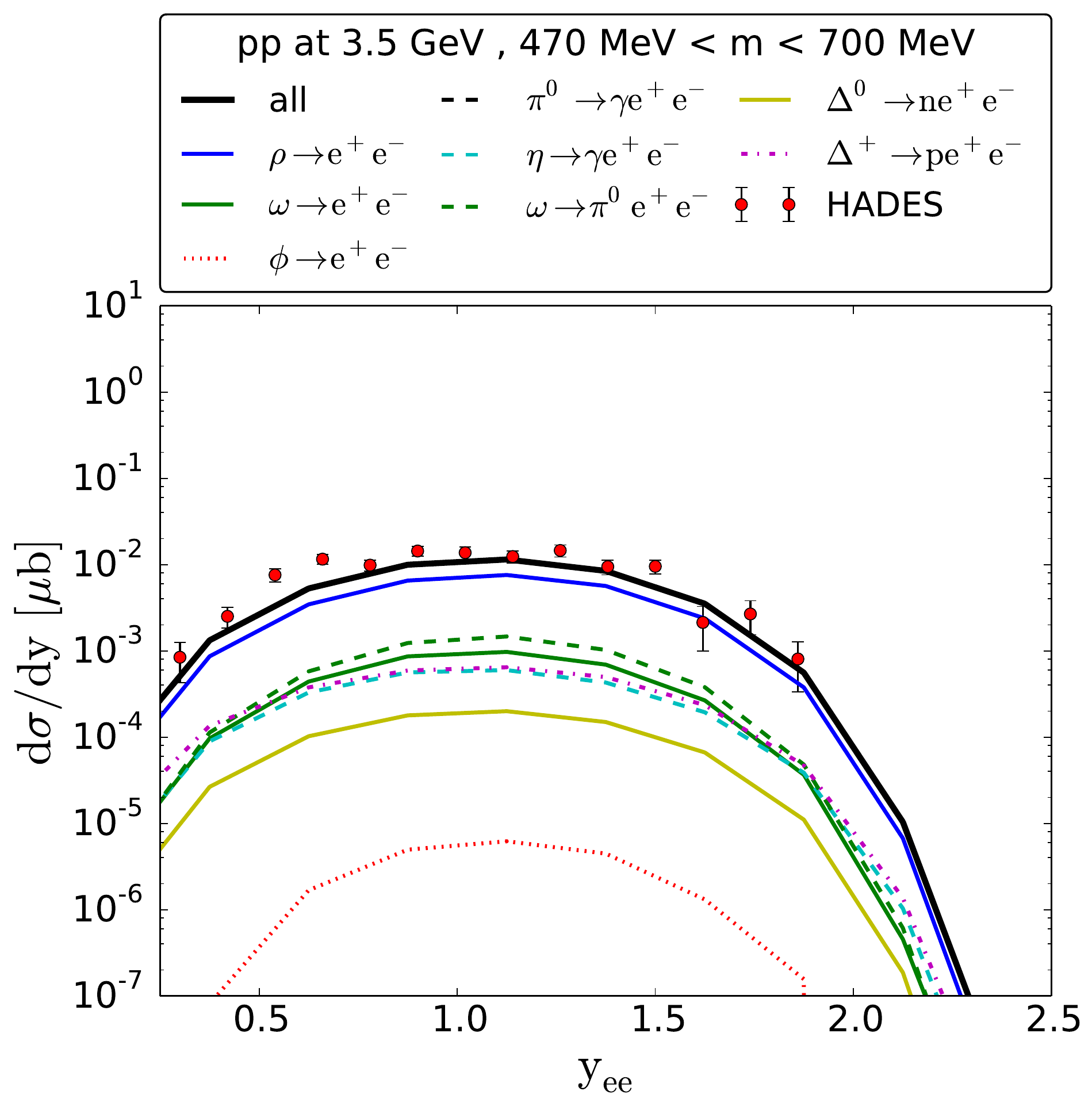}
    \includegraphics[width=0.45\textwidth]{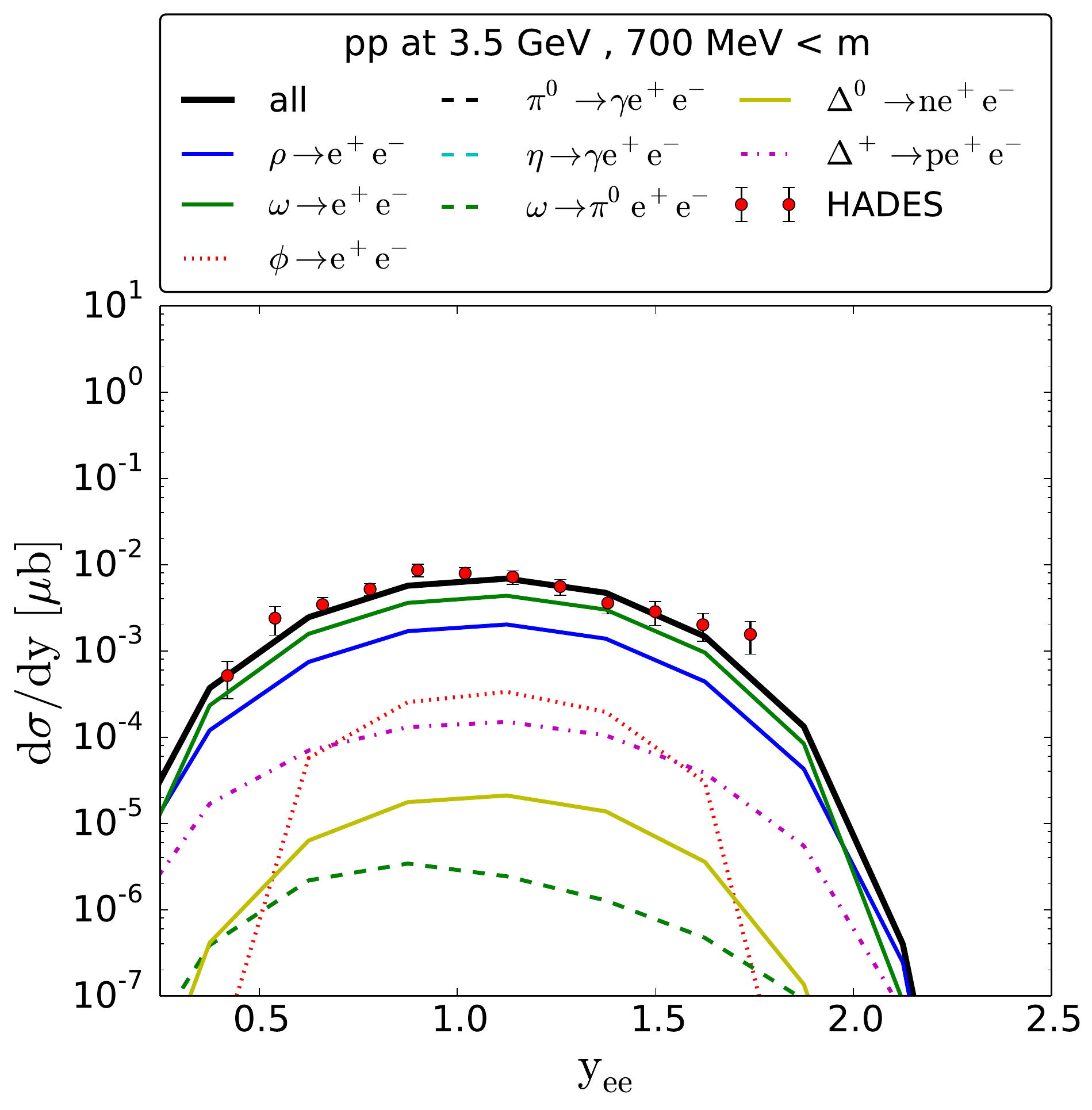}
    \caption{\label{fig:pp3.5ys}Rapidity spectra of dielectrons produced by pp collisions at $E_{\rm{Kin}}=3.5\,\textrm{GeV}$ in different invariant mass windows. Experimental data from \cite{HADES:2011ab}.}
\end{figure*}

The lepton pair transverse momentum and rapidity from reactions at $E_{\rm{Kin}}=3.5\,\textrm{GeV}$ are compared to experimental data from \cite{HADES:2011ab} in different invariant mass windows that reflect the different dominant contributions over the invariant mass range. Below $150$ MeV the $\pi^0$ Dalitz decay dominates. Between $150$ MeV and $470$ MeV the $\eta$ decay is the largest contribution, while above $470$ MeV and below $700$ MeV the $\rho$ channel exceeds the others. Above $700$ MeV the $\omega$ peak is dominant. The plots in Fig.~\ref{fig:pp3.5pts} and Fig.~\ref{fig:pp3.5ys} show that also for those more differential spectra that probe specific channels and different kinematic observables that probe different regions of the phase space agreement with experimental data is reasonable.

\bibliography{paper}

\end{document}